\documentclass[12pt,a4paper]{article}
\pdfoutput=1

\usepackage{color}
\usepackage{amssymb,amsmath,bm,bbm}
\usepackage{epsf}
\usepackage{epsfig}
\usepackage{afterpage}
\usepackage{longtable}
\usepackage[dvipsnames]{xcolor}
\usepackage[linktoc=page,bookmarks=false,colorlinks=false,
linkbordercolor=RoyalBlue,citebordercolor=ForestGreen,urlbordercolor=CornflowerBlue]{hyperref}
\usepackage{latexsym,mathrsfs,dsfont}
\usepackage[normalem]{ulem}
\usepackage[compress]{cite}
\usepackage{graphicx}
\usepackage{url}
\usepackage{paralist}
\usepackage{bbold}

\numberwithin{equation}{section}

\newcommand{\IM}{{\rm Im}}
\newcommand{\RE}{{\rm Re}}
\newcommand{\imlt}{\IM\lambda_t}

\newcommand{\tev}{\, {\rm TeV}}
\newcommand{\gev}{\, {\rm GeV}}
\newcommand{\mev}{\, {\rm MeV}}

\newcommand{\vcb}{|V_{cb}|}
\newcommand{\vtd}{|V_{td}|}
\newcommand{\vub}{|V_{ub}|}
\newcommand{\vts}{|V_{ts}|}
\newcommand{\vus}{|V_{us}|}

\newcommand{\bsi}{B_6^{(1/2)}}
\newcommand{\bei}{B_8^{(3/2)}}

\def\epe{\varepsilon'/\varepsilon}
\newcommand{\beq}{\begin{equation}}
\newcommand{\eeq}{\end{equation}}
\newcommand{\be}{\begin{equation}}
\newcommand{\ee}{\end{equation}}
\newcommand{\bi}{\begin{itemize}}
\newcommand{\ei}{\end{itemize}}
\newcommand{\ba}{\begin{array}}
\newcommand{\ea}{\end{array}}
\newcommand{\beqa}{\begin{eqnarray}}
\newcommand{\eeqa}{\end{eqnarray}}
\newcommand{\bea}{\begin{eqnarray}}
\newcommand{\eea}{\end{eqnarray}}
\newcommand{\beqn}{\begin{eqnarray}}
\newcommand{\eeqn}{\end{eqnarray}}

\newcommand{\D}{\Delta}

\newcommand{\eps}{\epsilon}

\newcommand{\B}{{\mathcal B}}

\definecolor{red}{cmyk}{0,1,1,0.4}

\def\Bqmumu{B_q\to \mu^+\mu^-}

\def\Bsmumu{B_s\to \mu^+\mu^-}
\def\kpn{K^+\rightarrow\pi^+\nu\bar\nu}
\def\klpn{K_{L}\rightarrow\pi^0\nu\bar\nu}

\def\D{\bf\color{blue}}

\usepackage{fancyhdr}
\advance \headheight by 3.0truept       

\begin{document}

\begin{flushright}
        {FLAVOUR(267104)-ERC-88}
\end{flushright}

\medskip

\renewcommand*{\thefootnote}{\fnsymbol{footnote}}
\begin{center}
{\Large\bf
\boldmath{$\kpn$ and $\klpn$ in the Standard Model:\\ Status and Perspectives}}
\\[0.8 cm]
{\bf Andrzej~J.~Buras,\footnote{\href{mailto:andrzej.buras@tum.de}{andrzej.buras@tum.de}} Dario Buttazzo,\footnote{\href{mailto:dario.buttazzo@tum.de}{dario.buttazzo@tum.de}} \\
Jennifer Girrbach-Noe,\footnote{\href{mailto:jennifer.girrbach@tum.de}{jennifer.girrbach@tum.de}} and Robert Knegjens\footnote{\href{mailto:robert.knegjens@tum.de}{robert.knegjens@tum.de}}
 \\[0.5 cm]}
{\small\it
TUM Institute for Advanced Study, Lichtenbergstr.~2a, D-85748 Garching, Germany\\
Physik Department, Technische Universit\"at M\"unchen,
James-Franck-Stra{\ss}e, \\D-85748 Garching, Germany}
\end{center}

\vskip1cm

\abstract{%
\noindent
In view 
of the recent start of the NA62 experiment at CERN that is expected to measure the $\kpn$ branching ratio with a precision of $10\%$, we summarise the present status of this promising decay within the Standard Model (SM).
We do likewise for the closely related $\klpn$, which will be measured by the KOTO experiment around 2020. 
As the perturbative QCD and electroweak corrections in both decays are under full control, the dominant uncertainties within the SM presently originate  from the CKM parameters $\vcb$, $\vub$ and $\gamma$. 
We show this dependence with the help of analytic expressions as well as accurate interpolating formulae.
Unfortunately a clarification of the discrepancies between inclusive and exclusive determinations of $\vcb$ and $\vub$ from tree-level decays will likely require results from the Belle II experiment available at the end of this decade. 
Thus we investigate whether higher precision on both branching ratios  is achievable by  determining $\vcb$, $\vub$ and $\gamma$ by means of other observables that are already precisely measured. 
In this context $\varepsilon_K$ and  $\Delta M_{s,d}$, together with the expected progress in QCD lattice calculations  will play a prominent 
role. We find $\mathcal{B}(\kpn)= \left(9.11\pm 0.72\right) \times 10^{-11}$ and $\mathcal{B}(\klpn)= \left(3.00\pm 0.30 \right) \times 10^{-11}$, which is more precise than using averages of the present tree-level values of 
$\vcb$, $\vub$ and $\gamma$.
Furthermore, we point out the correlation 
between $\mathcal{B}(\kpn)$, $\overline{\mathcal{B}}(B_s\to\mu^+\mu^-)$  and $\gamma$ within the SM, that is only very weakly dependent on  other CKM parameters.
Finally, we update the correlation of  $\klpn$ with the ratio $\epe$ in the SM taking the recent progress on $\epe$ from lattice QCD and the large $N$ approach into account.
}

\vfill

\thispagestyle{empty}
\newpage
\setcounter{page}{1}

\tableofcontents

\renewcommand*{\thefootnote}{\arabic{footnote}}
\setcounter{footnote}{0}


\section{Introduction}

The measurements of the branching ratios of the two {\it golden} modes
$\kpn$ and $\klpn$ will be among the top highlights of flavour physics in the rest of this decade. $\kpn$ is CP conserving while $\klpn$ is governed by 
CP violation. Both decays are dominated in the SM and in many of its
extensions by $Z$ penguin diagrams.
These decays are theoretically 
very clean, and the calculation of their branching ratios within the SM includes  next-to-leading order (NLO) QCD corrections to the top quark contributions 
\cite{Buchalla:1993bv,Misiak:1999yg,Buchalla:1998ba}, 
 NNLO QCD corrections to the charm contribution \cite{Buras:2005gr,Buras:2006gb,Gorbahn:2004my} and  NLO electroweak corrections \cite{Brod:2008ss,Brod:2010hi,Buchalla:1997kz} to both top and charm contributions.
Moreover, extensive calculations of isospin breaking effects and 
non-perturbative effects have been performed \cite{Isidori:2005xm,Mescia:2007kn}.
 Reviews of these two decays can be found in 
\cite{Buras:2004uu,Isidori:2006yx,Smith:2006qg,Komatsubara:2012pn,Buras:2013ooa,Blanke:2013goa,Smith:2014mla} and their power in probing energy scales as high as several hundreds of $\tev$ has been demonstrated in \cite{Buras:2014zga}.

In view of the recent start of the NA62 experiment at CERN that is expected to measure the $\kpn$ branching ratio with a precision of $10\%$  compared to the SM prediction
\cite{Rinella:2014wfa,Romano:2014xda}, and the expected 
measurement of $\klpn$  by KOTO around 2020 at J-PARC \cite{Komatsubara:2012pn,Shiomi:2014sfa}, it is the right time to summarise the present  status of these decays  within the SM. This is 
motivated in particular by the fact that different estimates appear in the 
literature due to different inputs used for the Cabibbo-Kobayashi-Maskawa (CKM) matrix elements, which presently 
constitute the main uncertainty in the SM predictions for these two branching 
ratios. This has been stressed in \cite{Buras:2014sba}, where 
the dependence of both branching ratios on the chosen values of  
$\vcb$ and $\vub$ extracted from tree-level decays has been studied (see Table 3 of that paper).

At this point two strategies for the determination of the contribution of 
the SM dynamics to these decays are envisaged:
\paragraph{Strategy A:}
The CKM matrix is determined using tree-level measurements of 
\be\label{STRA}
\vus,\qquad \vcb, \qquad \vub, \qquad \gamma,
\ee
where $\gamma$ is an angle of the unitarity triangle (UT). As New Physics (NP)
seems to now be well separated from the electroweak scale, this determination of the CKM matrix is not expected to be
polluted by NP contributions.\footnote{Recent analyses of  the room left for NP in tree-level decays can be found in \cite{Bobeth:2014rda,Bobeth:2014rra,Brod:2014bfa}.} Inserting these inputs
into the known expressions for the relevant branching ratios (see Section~\ref{sec:2}) then allows a determination of the SM values for the $K\to \pi \nu\bar \nu$ branching ratios independently of whether NP is present at short distance scales or not. The departure of these predictions from future data would therefore
allow us to discover whether NP contributes to these decays independently of 
whether it contributes to other decays or not. This information is clearly 
important for the selection of  successful extensions of the SM through 
flavour-violating processes.

Unfortunately, this strategy cannot be executed in a satisfactory manner 
at present due to  the discrepancies between inclusive and exclusive determinations of $\vcb$ and $\vub$ from tree-level decays. Moreover, the precision on $\gamma$ from tree-level decays 
is still unsatisfactory for this purpose. While the measurement of $\gamma$ 
should be significantly improved by  LHCb in the coming years, 
discrepancies between inclusive and exclusive determinations of $\vcb$ and $\vub$ from tree-level decays are likely to be resolved only by the time of the Belle II experiment at SuperKEKB  at the end of this decade. 

The clarification of the discrepancies between inclusive and exclusive determinations of $\vcb$ and $\vub$ from tree-level decays is important, but there are reasons to
expect that the exclusive determinations will eventually be the ones to be favoured. First of all, exclusive measurements are easier to perform than the inclusive ones. Equally important, due to the significant improvement in 
the calculations of the relevant form factors by lattice QCD, exclusive determinations are more straightforward than the inclusive ones. This is opposite to the philosophy of ten years ago, where QCD lattice calculations were still at the early stage and inclusive determinations were favoured.

Yet, from the present perspective it is useful to study the SM predictions for 
$\kpn$ and $\klpn$ in the full range of $\vcb$, $\vub$ and $\gamma$ known 
from tree-level decays, as this will clearly demonstrate the need for the 
reduction of parametric uncertainties. This will also allow the SM predictions for these decays to be monitored as the determination of $\gamma$ will improve in the coming years at the LHC. This should be of interest in view of the first 
results on $\kpn$ from NA62, which are expected already in 2016. Moreover, it will also
be of interest to see how other observables, like $\varepsilon_K$, 
$\Delta M_s$, $\Delta M_d$, and rare $B_{s,d}$ decays are modified 
when the parameters in (\ref{STRA}) are varied, and what their correlations 
with $\kpn$ and $\klpn$ are within the SM. As we will see, some of these correlations are practically independent of $\vcb$ and $\vub$ and as such are particularly suited for a precise tests of the SM.

\paragraph{Strategy B:} 
Here the assumption is made that the SM is the whole story and 
all available information from flavour-changing neutral current (FCNC) processes is used to determine the 
CKM matrix. Our strategy here will be to ignore tree-level determinations 
of $\vub$ and $\vcb$, as the discrepancies mentioned above could also result 
from experimental data, which will improve only at the end of this decade.
Similarly, the tree-level determination of $\gamma$ will be
left out.
Then the observables to be used for the determination of the CKM parameters will be\,\footnote{Note that the present determination of $S_{\psi\phi}$ has no impact on the CKM parameters in the SM.}
\be\label{STRB}
\varepsilon_K, \qquad \Delta M_s, \qquad \Delta M_d, \qquad S_{\psi K_S},
\ee
accompanied by lattice QCD calculations of the relevant non-perturbative parameters. In this manner also $\vcb$, $\vub$ and $\gamma$ can be determined. This 
is basically what the UTfit \cite{Bona:2006ah} and CKMfitter \cite{Charles:2015gya} collaborations do, except that we 
ignore the tree-level determinations of $\vcb$, $\vub$ and $\gamma$ for the reasons stated above. 
As the dominant top quark contribution to $\varepsilon_K$ is proportional to $\vcb^4$ and $\Delta M_{s,d}$ are proportional to $\vcb^2$, a useful determination of $\vcb$ can be obtained from these quantities.\footnote{The strategy 
for the determination of $\vcb$ from $\varepsilon_K$ is not new \cite{Kettell:2002ep} and has been considered recently in \cite{Buras:2013raa}. See in particular formula (29) in that paper.} The full UT
is then constructed by using the ratio $\Delta M_d/\Delta M_s$ and $S_{\psi K_S}$.  We find that with the most recent lattice 
QCD input on the parameter $\xi$ \cite{Bouchard:2014eea}, the determination of $\gamma$ in this manner is impressive,  and  also the value of $\vcb$ is significantly more accurate than 
from tree-level decays. In the case of $\vub$  the accuracy is found to be comparable 
to the most recent exclusive determination \cite{Bailey:2014fpx}.

It should be emphasised that while strategy A is ultimately the one to use to study extensions of the SM, the virtue of strategy B at present is the greater accuracy of the SM predictions for the observables that we consider. 
By simply imposing constraints from several measurements  we arrive at narrow ranges for the parameters in (\ref{STRA}) -- given that the SM is the whole story.

\bigskip

In the present paper we will follow these two strategies using the most 
recent inputs relevant for both of them, in particular the  ones
from lattice QCD. In Section~\ref{sec:2} we summarise 
the present status of the $\kpn$ and $\klpn$ decays in the SM and discuss the main 
uncertainties with the help of analytic expressions.
In Sections~\ref{sec:A} and \ref{sec:B} we follow strategies A and B, respectively, and present in some detail our numerical results.  In Section~\ref{sec:4a} we present an updated analysis of  the correlation of $\klpn$ and the ratio $\epe$ in the SM.
We conclude in Section~\ref{sec:concl}. In the appendices we collect a number of 
additional expressions that we used in our analysis.


\section{Basic formulae}\label{sec:2}

We present here the basic formulae for the branching ratios for the $\kpn$ and $\klpn$ decays in the SM.
This section can be considered as an update to the analogous section (Section 2) of \cite{Buras:2004uu}, a review of these decays from 2007.
The main advances in the last eight years are:
\begin{itemize}
\item
computation of complete NLO electroweak corrections to the charm quark contribution to $\kpn$ in \cite{Brod:2008ss};
\item
computation of complete NLO electroweak corrections to the top quark contribution 
to $\kpn$ and $\klpn$ in \cite{Brod:2010hi};
\item
reduction of uncertainties due to $m_t(m_t)$, $m_c(m_c)$ and $\alpha_s(M_Z)$, with the last two relevant 
in particular for the charm contribution to $\kpn$.
\end{itemize}

While incorporating these advances in our presentation we will also include 
\begin{itemize}
\item
NLO QCD corrections to the top quark contributions 
\cite{Buchalla:1993bv,Misiak:1999yg,Buchalla:1998ba} and NNLO QCD corrections to the
charm contribution \cite{Buras:2005gr,Buras:2006gb,Gorbahn:2004my};
\item
isospin breaking effects and 
non-perturbative effects\cite{Isidori:2005xm,Mescia:2007kn}.
\end{itemize}


\boldmath
\subsection{$\kpn$}
\unboldmath

The branching ratio for $\kpn$ in the SM is dominated
by $Z^0$ penguin diagrams, with a significant contribution from box 
diagrams. Summing over three neutrino flavours, it can be written
as follows \cite{Buchalla:1998ba, Mescia:2007kn}  
\begin{align}\label{bkpnn}
\mathcal{B}(K^+\to\pi^+\nu\bar\nu) = \kappa_+ (1+\Delta_\text{EM})\cdot
&\left[\left(\frac{{\rm Im}\lambda_t}{\lambda^5}X(x_t)\right)^2\right. \notag\\
&\left. +\left(\frac{{\rm Re}\lambda_c}{\lambda}P_c(X)+
\frac{{\rm Re}\lambda_t}{\lambda^5}X(x_t)\right)^2\right],
\end{align}
with
\begin{equation}\label{kapp}
\kappa_+ = (5.173\pm 0.025 )\cdot 10^{-11}\left[\frac{\lambda}{0.225}\right]^8, \qquad\qquad \Delta_\text{EM} = -0.003.
\end{equation}
Here $x_t=m^2_t/M^2_W$, $\lambda=\vus$, $\lambda_i=V^*_{is}V_{id}$ are the CKM factors
discussed below, and $\kappa_+$ summarises the remaining factors, in particular the relevant hadronic matrix elements that can
be extracted from leading semi-leptonic decays of $K^+$, $K_L$ and $K_S$ mesons
\cite{Mescia:2007kn}. $\Delta_{\rm EM}$ describes the electromagnetic radiative correction from photon exchanges. $X(m_t)$ and $P_c(X)$ are the loop functions for the top and charm quark contributions, which are discussed below.
An explicit derivation of (\ref{bkpnn}) can be found in \cite{Buras:1998raa}.
The apparent large sensitivity of $\mathcal{B}(\kpn)$ to $\lambda$ is spurious as 
$P_c(X)\sim \lambda^{-4}$ (see (\ref{p0k})) and the dependence on $\lambda$ in (\ref{kapp}) 
cancels the one in (\ref{bkpnn}) to a large extent.  
Therefore when changing 
$\lambda$ it is essential to keep track of  all the $\lambda$ dependence.

In obtaining the numerical values in~(\ref{kapp})~\cite{Mescia:2007kn}, the $\overline{\text{MS}}$ scheme with
\be\label{INPUT}
\sin^2\theta_w(M_Z)=0.23116, \qquad\qquad \alpha(M_Z)=\frac{1}{127.925}, 
\ee
has been used.
As their errors are below $0.1\%$ these can currently be neglected.
Note, however, that although the prefactor of the effective Hamiltonian, $\alpha/\sin^2\theta_w$, is precisely
known in a particular renormalisation scheme ($\overline{\text{MS}}$ in this case) it remains a scheme dependent quantity,
with the scheme dependence only removed by considering higher order 
electroweak effects in $K\to \pi \nu\bar\nu$.
An analysis of such effects in the large $m_t$ limit \cite{Buchalla:1997kz} demonstrated that in principle this scheme dependence  could introduce a $\pm 5\%$ correction in the $K\to \pi \nu\bar\nu$ branching ratios, and that with the $\overline{\text{MS}}$ definition of $\sin^2\theta_W$ these higher order 
electroweak corrections are found below $2\%$. 
  However, only the complete  analysis of two-loop
electroweak contributions to $K\to \pi \bar \nu \nu$  in \cite{Brod:2010hi} for 
the top contribution could put such expectations on  firm footing. 
The same applies to the NLO electroweak effects in the charm contribution to 
$\kpn$ evaluated in \cite{Brod:2008ss}.

The short distance function $X(x_t)$ relevant for the top quark contribution,
 including NLO QCD corrections \cite{Buchalla:1993bv,Misiak:1999yg,Buchalla:1998ba} and two-loop electroweak contributions \cite{Brod:2010hi}, is
\be\label{XSM}
X(x_t)=1.481\pm0.005_{\rm th}\pm 0.008_{\rm exp},
\ee
where the first error comes from the 
remaining renormalisation scale and scheme uncertainties, as well as the theoretical error on the $\overline{\text{MS}}$ parameters due to the matching at the electroweak scale, while the second one corresponds to the combined experimental error on the top and 
$W$ masses entering the ratio $x_t$, and on the strong coupling $\alpha_s(M_Z)$. The central value and errors in \eqref{XSM} have been obtained using the $\overline{\text{MS}}$ couplings with full NNLO precision \cite{Buttazzo:2013uya} -- 3-loop 
running in the SM and 2-loop matching at the weak scale (plus 4-loop QCD running of $\alpha_s$ and 3-loop QCD matching in $\alpha_s$ and $y_t$) -- and varying the renormalisation scale between $M_t/2$ and $2M_t$. The NLO EW correction has been included,
using the result presented in \cite{Brod:2010hi}, in order to eliminate the large EW renormalisation scheme dependence of 
the pure QCD result. See Appendix~\ref{app:X} for details about the different contributions to $X(x_t)$.

The parameter $P_c(X)$ summarises the charm contribution
and is defined through
\be\label{PCX}
P_c(X)=P_c^{\rm SD}(X)+\delta P_{c,u}, \qquad \delta P_{c,u}=0.04 \pm 0.02,
\ee
with the long-distance contributions $\delta P_{c,u}$ calculated in
\cite{Isidori:2005xm}. Future lattice calculations could reduce the present 
error in this part \cite{Isidori:2005tv}.
The short-distance part is given by
\begin{equation}\label{p0k}
P_c^{\rm SD}(X)=\frac{1}{\lambda^4}\left[\frac{2}{3} X^e_{\rm NNL}+\frac{1}{3}
 X^\tau_{\rm NNL}\right]
\end{equation}
where the functions
$X^\ell_{\rm NNL}$ result from QCD  NLO  \cite{Buchalla:1998ba,Buchalla:1993wq} and NNLO calculations\cite{Buras:2005gr,Buras:2006gb}. They also include complete 
two-loop electroweak contributions \cite{Brod:2008ss}.
The index ``$\ell$" distinguishes between the charged lepton flavours in 
the box diagrams. This distinction is irrelevant in the top contribution 
due to $m_t\gg m_\ell$ but is relevant in the charm contribution as 
$m_{\tau}>m_c$.
The inclusion of NLO and NNLO  QCD corrections have
reduced considerably the large dependence on the renormalisation scale $\mu_c$ (with $\mu_c={\cal O}(m_c)$) present in the leading order expressions for the charm contribution. 
The two-loop electroweak corrections on the other hand reduced the dependence on the definition of electroweak parameters.
An excellent approximation for $P_c^{\rm SD}(X)$, including all these corrections,  as a function of $\alpha_s(M_Z)$ and $m_c(m_c)$ is given in (50) 
of \cite{Brod:2008ss} (see Appendix~\ref{app:Pc}). 
Using this formula for the most recent input parameters 
\cite{Chetyrkin:2009fv,Beringer:1900zz}
\be\label{recentInputs}
\lambda=0.2252(9), \qquad m_c(m_c)=1.279(13)\gev, \qquad \alpha_s(M_Z)=0.1185(6)
\ee
we find
\be
P^\text{SD}_c(X)= 0.365\pm 0.012.
\ee
Adding the long distance contribution in (\ref{PCX}) we finally find
\be\label{PCFINAL}
P_c(X)= 0.404\pm 0.024,
\ee
where we have added the errors in quadratures.
We  will use this value in our numerical analysis. In obtaining the error in (\ref{PCFINAL}) we kept $\lambda$ fixed at its central value, as its error is very small and the strong dependence 
on $\lambda$ in $P^\text{SD}_c(X)$ is canceled by 
other factors in the formula for the branching ratio as discussed above. 


\boldmath
\subsection{$\klpn$}
\unboldmath

The branching ratio for $\klpn$ in the SM is  fully dominated by the 
diagrams with internal top exchanges,
with the charm contribution well below $1\%$. It can be written then
as follows \cite{Buchalla:1996fp,Buchalla:1995vs}
\begin{equation}\label{bklpn}
\mathcal{B}(\klpn)=\kappa_L\cdot
\left(\frac{{\rm Im}\lambda_t}{\lambda^5}X(x_t)\right)^2,
\end{equation}
where \cite{Mescia:2007kn}
\begin{equation}\label{kapl}
\kappa_L=
(2.231\pm 0.013)\cdot 10^{-10}\left[\frac{\lambda}{0.225}\right]^8.
\end{equation}
We have summed over three neutrino flavours. 
An explicit derivation of (\ref{bklpn}) can be found 
in \cite{Buras:1998raa}.
Due to the absence of $P_c(X)$ in (\ref{bklpn}), the theoretical uncertainties in
$\mathcal{B}(K_L\to\pi^0\nu\bar\nu)$ are due only to $X(x_t)$ and amount to about $1\%$ at the level of the branching ratio.
The main uncertainty then comes from $\imlt$, which is by far dominant with respect to the other parametric uncertainties due to $\kappa_L$ and  $m_t$, with the latter present in $X(x_t)$.

\subsection{Experimental prospects}
 Experimentally we have \cite{Artamonov:2008qb}
\be\label{EXP1}
\mathcal{B}(\kpn)_\text{exp}=(17.3^{+11.5}_{-10.5})\cdot 10^{-11}\,,
\ee
and the $90\%$ C.L. upper bound \cite{Ahn:2009gb}
\be\label{EXP2}
\mathcal{B}(\klpn)_\text{exp}\le 2.6\cdot 10^{-8}\,.
\ee

The prospects for improved measurements of $\mathcal{B}(\kpn)$ are very good.
One should stress that already a measurement of this  branching ratio with an
accuracy of $10\%$ will give us a very important insight into the physics 
at short distance scales. Indeed the NA62 experiment at CERN \cite{Rinella:2014wfa,Romano:2014xda} is aiming to reach this precision, and it is expected to accumulate 100 SM events with a good signal over background figure by 2018. 
 In order to achieve a $5\%$ measurement of the branching ratio, which will be the next goal of NA62, more time is needed. The planned new experiment at Fermilab (ORKA) could in principle reach the accuracy of $5\%$ \cite{E.T.WorcesterfortheORKA:2013cya}.\footnote{Unfortunately the US P5 committee did not recommend moving ahead with ORKA and it appears that the precision on   $\mathcal{B}(\kpn)$ will depend in the coming ten years entirely on the progress made by NA62.}

Concerning $\klpn$, the KOTO experiment at  J-PARC  aims in the first step in 
measuring  $\mathcal{B}(\klpn)$ at SM sensitivity and 
should provide interesting results around 2020 on this 
branching ratio \cite{Komatsubara:2012pn,Shiomi:2014sfa}. There are also plans to measure 
this decay at CERN and one should hope that Fermilab will contribute to 
these efforts in the next decade. 
The combination of $\kpn$ and $\klpn$  is particularly powerful in testing NP. Assuming that 
NA62 and KOTO will reach the expected precision and the branching ratios on 
these decays will be at least as high as the ones predicted in the SM, these 
two decays are expected to be the superstars of flavour physics after 2018.


\section{CKM inputs from tree-level observables}\label{sec:A}

\begin{table}[!tb]
\renewcommand{\arraystretch}{1.2}
\centering%
\begin{tabular}{|cl|cl|}
\hline\hline
$|\eps_K|$ & $2.228(11)\times 10^{-3}$\hfill\cite{Beringer:1900zz} 
&
$F_K $ & $ 156.1(11)\mev$\hfill\cite{Aoki:2013ldr}
\\
$S_{\psi K_{\rm S}}$ & $0.682(19)$\hfill\cite{Amhis:2012bh} 
&
$\hat{B}_K$ & $0.750(15)$ \hfill \cite{Aoki:2013ldr,Buras:2014maa}
\\
$\Delta M_K$ & $ 0.5292(9)\times 10^{-2} \,\text{ps}^{-1}$\hfill\cite{Beringer:1900zz} 
&
$F_{B_d}$ & $190.5(42)\mev$ \hfill \cite{Aoki:2013ldr} 
\\
$\Delta M_d $ & $ 0.507(4)\,\text{ps}^{-1}$\hfill\cite{Amhis:2012bh} 
& 
$F_{B_s}$ & $227.7(45)\mev$ \hfill \cite{Aoki:2013ldr} 
\\
$\Delta M_s $ & $ 17.761(22)\,\text{ps}^{-1}$\hfill\cite{Amhis:2012bh}	 
& 
$F_{B_s}\sqrt{\hat{B}_{B_s}}$ & $266(18)\mev$\hfill\cite{Aoki:2013ldr}
\\
$\gamma$ & $\left(73.2^{+6.3}_{-7.0}\right)^\circ $\hfill\cite{Trabelsi:2014}	 
& 
$\xi $ & $1.268(63)$\hfill\cite{Aoki:2013ldr}
\\
$|V_{us}|$ & $0.2252(9)$\hfill\cite{Amhis:2012bh} 
& 
&
\\
\cline{3-4}
$\Delta\Gamma_s/\Gamma_s$ & $0.138(12)$\hfill\cite{Amhis:2012bh} 
& 
$\eta_{cc}$ & $1.87(76)$\hfill\cite{Brod:2011ty}
\\
$\tau_{B_d}$ & $ 1.519(5) \,\text{ps}$\hfill\cite{Amhis:2012bh} 
&
$\eta_{ct}$ & $ 0.496(47)$\hfill\cite{Brod:2010mj}
\\
$\tau_{B_s}$ & $ 1.512(7)\,{\rm ps}$\hfill\cite{Amhis:2012bh}
&
$\eta_{tt}$ & $0.5765(65)$\hfill\cite{Buras:1990fn}
\\
\cline{1-2}
$\alpha_s(M_Z)$ & $0.1185(6) $\hfill\cite{Beringer:1900zz}
&
$\eta_B$ & $0.55(1)$\hfill\cite{Buras:1990fn,Urban:1997gw}
\\
$m_c(m_c) $ & $ 1.279(13) \gev$ \hfill\cite{Chetyrkin:2009fv}
&
& 
\\
$M_t $ & $ 173.34(82)\gev$\hfill\cite{ATLAS:2014wva}
&
& 
\\
\hline\hline
\end{tabular} 
\caption{\it Values of theoretical and experimental quantities used as input parameters.}\label{tab:input}
\end{table}


\subsection{Determination of the branching ratios}

As discussed in the introduction, the CKM matrix can be determined by the tree-level measurements $\vub$, $\vcb$, $\vus$, and the angle $\gamma$ of the UT.
Although this is in principle the optimal strategy, it is currently marred by disagreements between the exclusive and inclusive determinations of both $\vub$ and $\vcb$ -- for recent reviews see \cite{Ricciardi:2014aya,Ricciardi:2013cda,Gambino:2015ima}.
We proceed to present the latest results of both determinations, as well as our weighted average, with which we will give the SM predictions in what we call strategy A.

The most recent exclusive determinations from lattice QCD form factors are~\cite{Bailey:2014tva, Bailey:2014fpx,Aoki:2013ldr}
\be\label{exclusive}
\vub_{\rm excl} =(3.72\pm0.14)\times 10^{-3}, \qquad  \vcb_{\rm excl}=(39.36\pm0.75)\times 10^{-3}.
\ee
The inclusive values are given by~\cite{Alberti:2014yda,Aoki:2013ldr}
\be\label{inclusive}
\vub_{\rm incl} =(4.40\pm0.25)\times 10^{-3}, \qquad  \vcb_{\rm incl}=(42.21\pm0.78)\times 10^{-3}.
\ee
We take a weighted average and scale the errors based on the resulting $\chi^2$ (specifically, we follow the method advocated in \cite{Beringer:1900zz}), which gives
\be\label{average}
\vub_{\rm avg} =(3.88\pm0.29)\times 10^{-3}, \qquad  \vcb_{\rm avg}=(40.7\pm1.4)\times 10^{-3}.
\ee
For the CKM angle $\gamma$ we take the current world average of direct measurements~\cite{Trabelsi:2014}
\begin{equation}\label{gamma}
    \gamma = (73.2^{+6.3}_{-7.0})^\circ.
\end{equation}
Using this, together with $|V_{us}|=\lambda$ already given in \eqref{recentInputs},
we can determine the full CKM matrix.

\begin{figure}[t]
\centering%
\begin{tabular}{ccccc}
{\small $\B(\kpn)$} &&&& {\small $\B(\klpn)$}\\[2pt]
\includegraphics[width=0.39\textwidth]{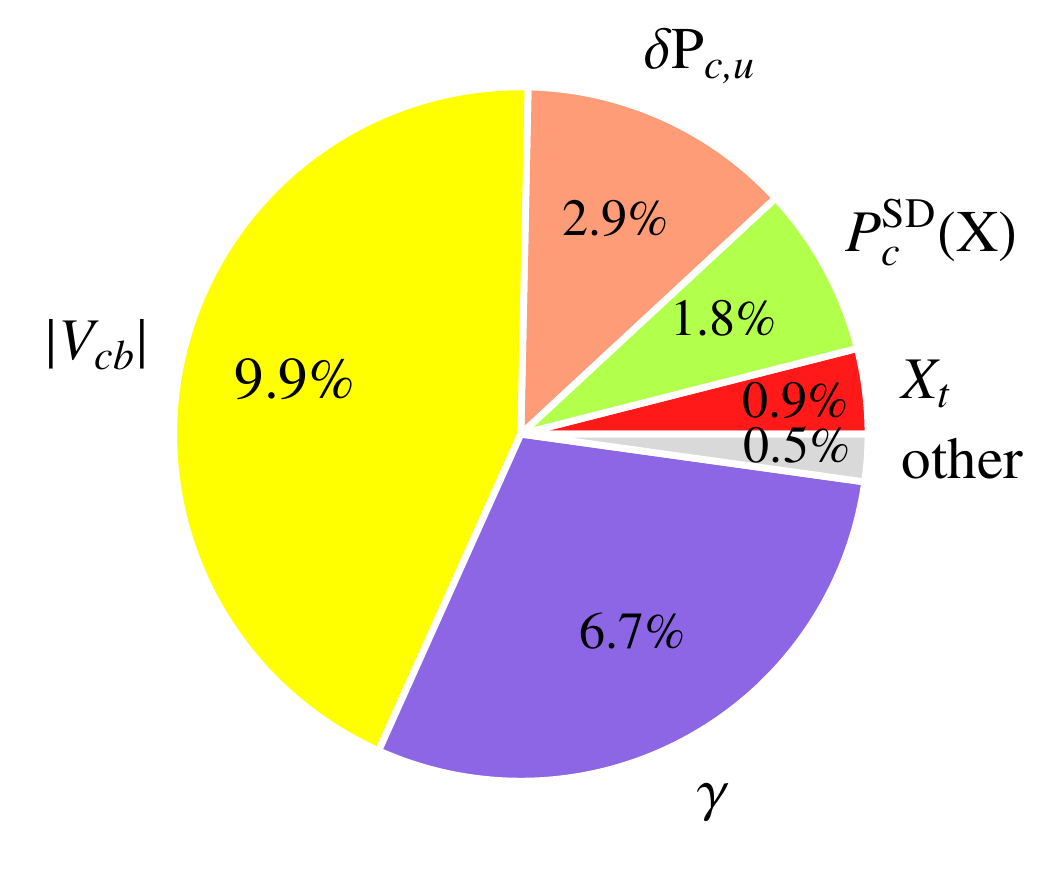} &&&&
\includegraphics[width=0.39\textwidth]{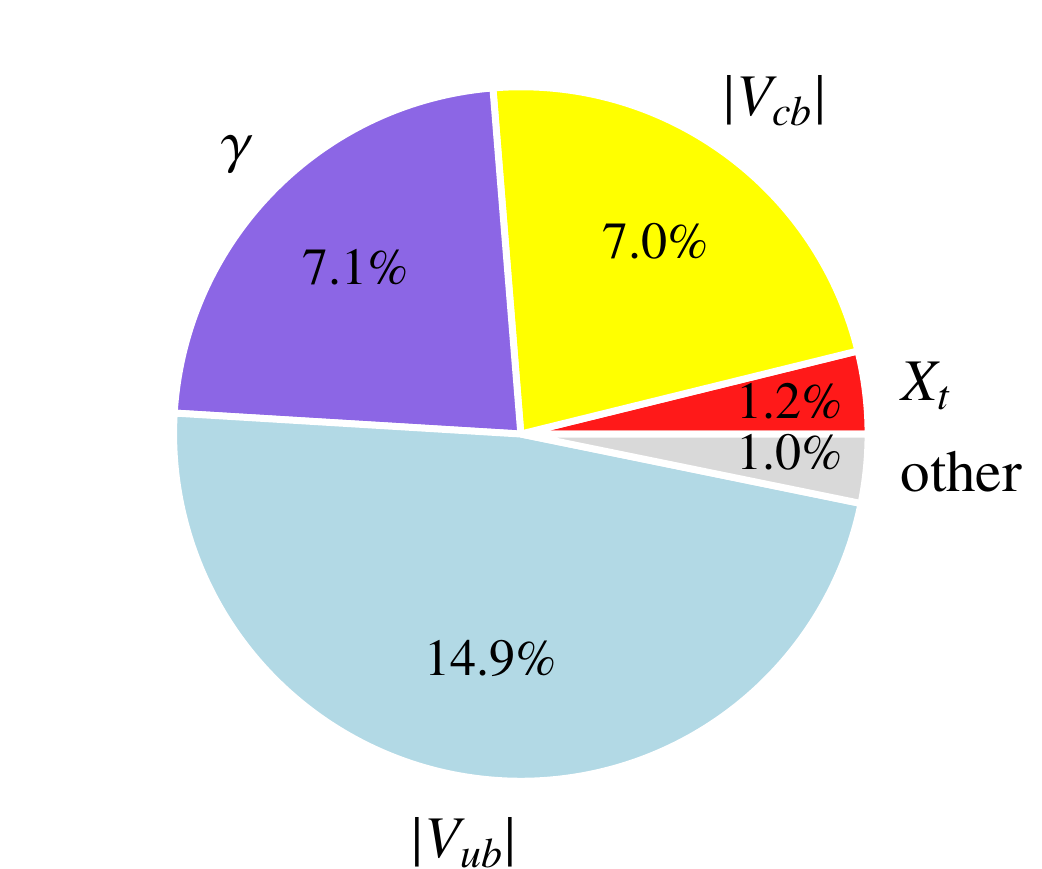}
\end{tabular}
\caption{\it Error budgets for the branching ratio observables $\mathcal{B}(\kpn)$ and $\mathcal{B}(\klpn)$. 
The remaining parameters, which each contribute an error of less than 1\%, are grouped into the ``other'' category.
\label{fig:err_budgets}}
\end{figure}

In particular, we can determine the quantities $\lambda_t$ and $\lambda_{c}$, which enter the expressions for the branching ratios in \eqref{bkpnn} and \eqref{bklpn}, as functions of these input parameters.
These expressions are:
\begin{align}
\RE\lambda_t &\simeq |V_{ub}||V_{cb}|\cos\gamma(1 - 2\lambda^2) + (|V_{ub}|^2 - |V_{cb}|^2)\lambda\left(1 - \frac{\lambda^2}{2}\right),\\
\IM\lambda_t &\simeq |V_{ub}||V_{cb}|\sin\gamma,\\
\RE\lambda_c &\simeq -\lambda\left(1 - \frac{\lambda^2}{2}\right),
\end{align}
which, with respect to their leading order in $\lambda$, are accurate up to $\mathcal{O}(\lambda^4)$ corrections.
The (exact) numerical values for $\RE\lambda_t$ and $\IM\lambda_t$ obtained from our three different choices of $V_{ub}$ and $V_{cb}$ in \eqref{exclusive}-\eqref{average} are given in Table~\ref{tab:SA}.

These expressions can then be directly inserted into \eqref{bkpnn} and \eqref{bklpn} in order to determine the two branching ratios.
Using our averages from \eqref{average} together with \eqref{gamma} gives
\begin{align}
    \mathcal{B}(\kpn) &= \left(8.4 \pm 1.0\right) \times 10^{-11}, \\
    \mathcal{B}(\klpn) &= \left(3.4 \pm 0.6\right) \times 10^{-11}.    
\end{align}
In Figure~\ref{fig:err_budgets} we show the error budgets for these two observables, and see that the CKM uncertainties dominate. 
In particular in the case of $\kpn$ we observe large uncertainties due to $\vcb$ and $\gamma$, while in the case of $\klpn$ the 
uncertainty due to $\vub$ dominates but the ones from $\vcb$ and $\gamma$ are also large. 
The remaining parameters, which each contribute an error of less than 1\%, are grouped into the ``other'' category.

\begin{figure}[t!]
\centering%
\includegraphics[width=0.48\textwidth]{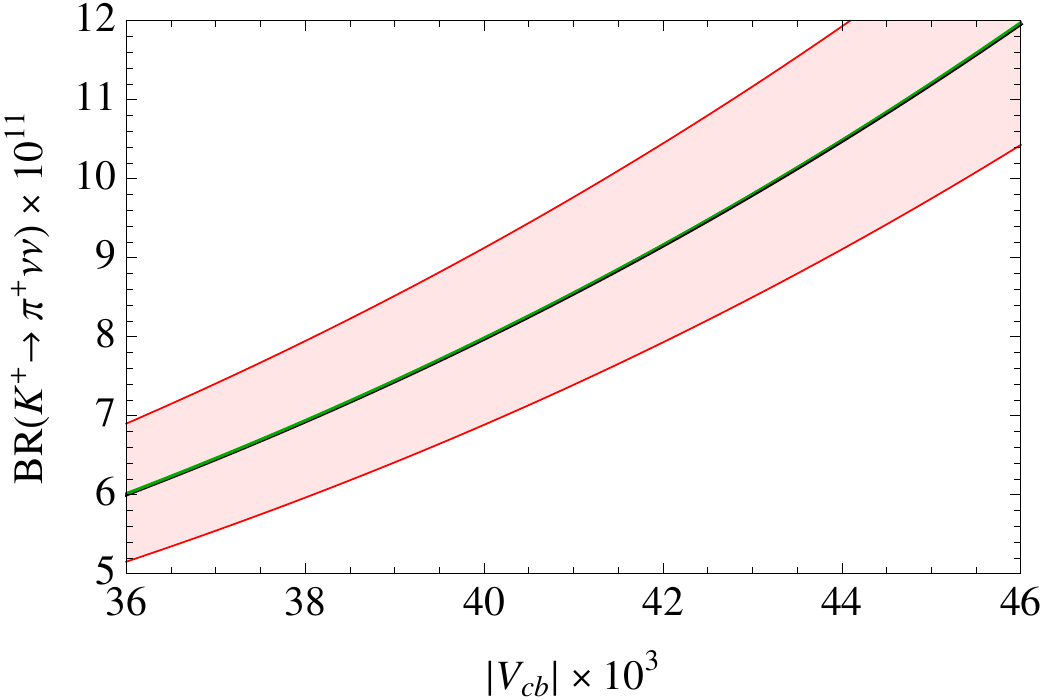}%
\hfill%
\includegraphics[width=0.48\textwidth]{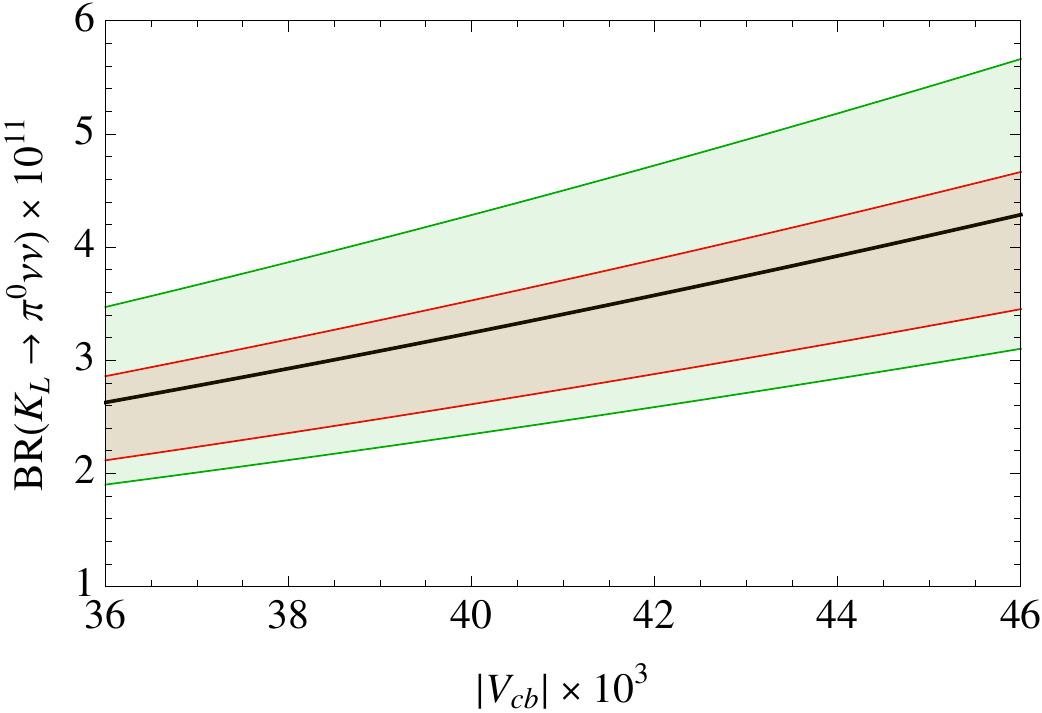}\\
\includegraphics[width=0.48\textwidth]{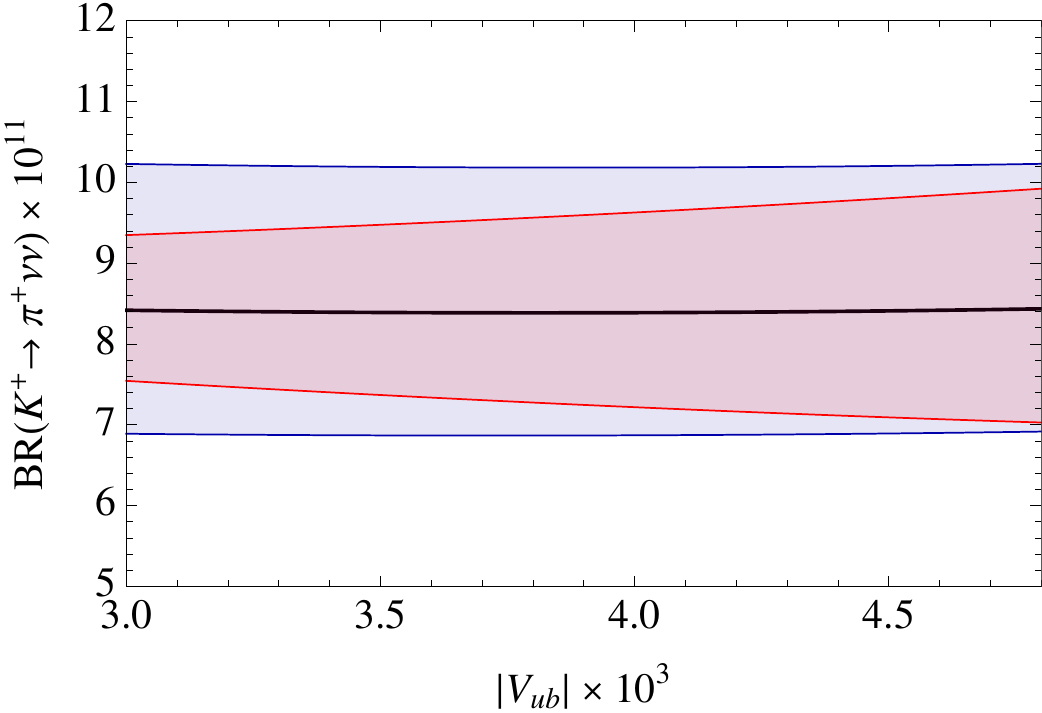}%
\hfill%
\includegraphics[width=0.48\textwidth]{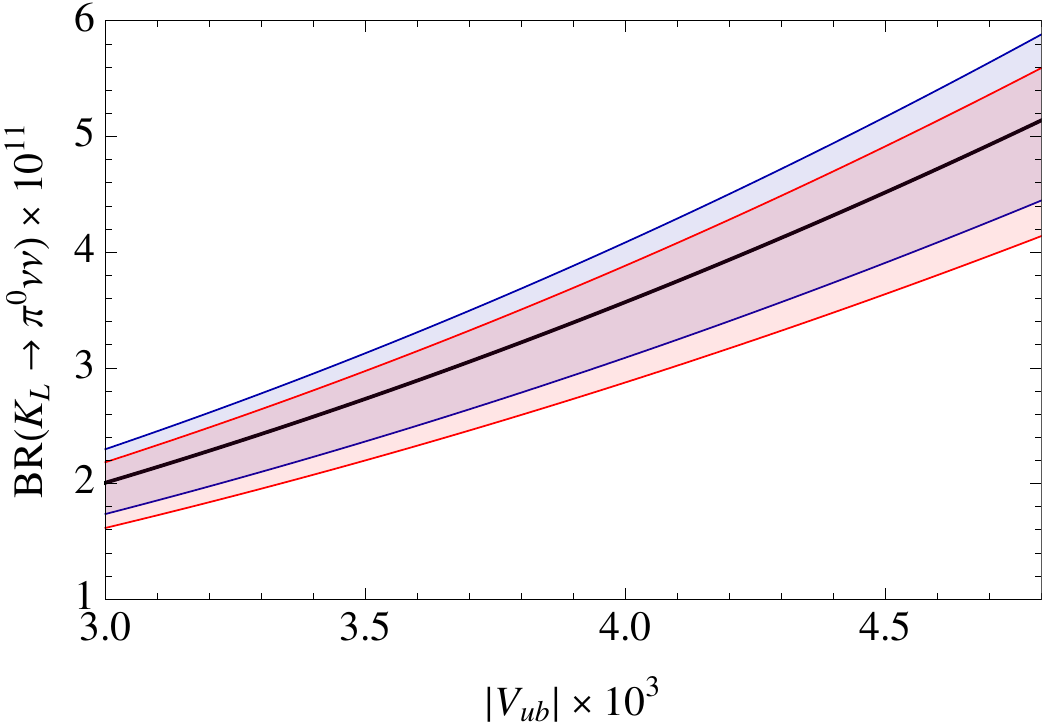}\\
\includegraphics[width=0.48\textwidth]{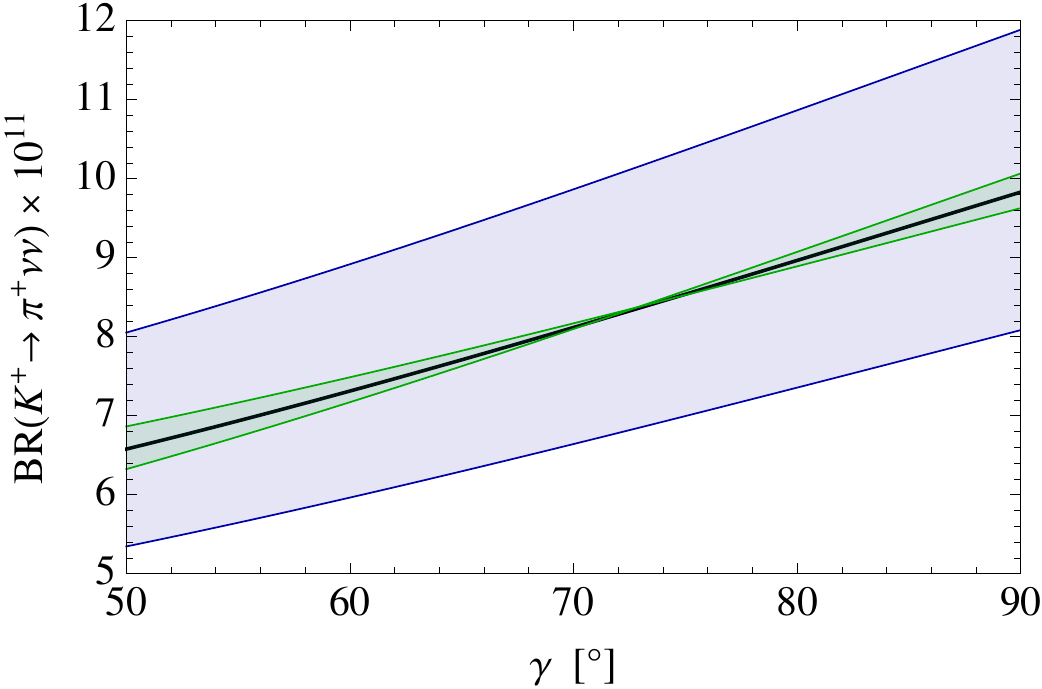}%
\hfill%
\includegraphics[width=0.48\textwidth]{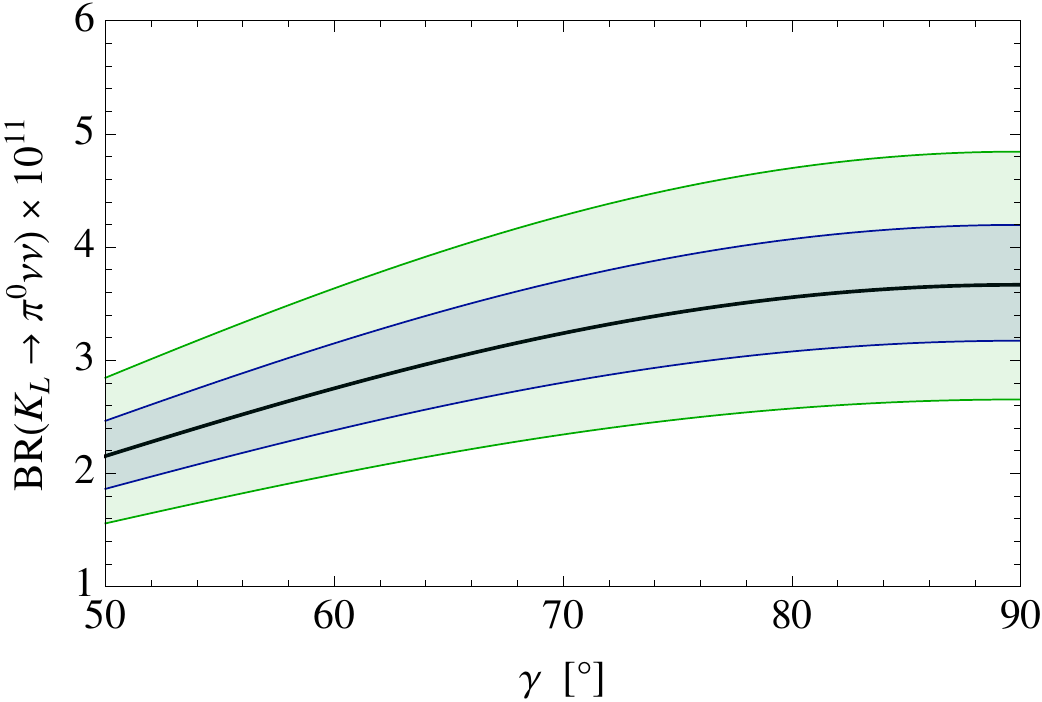}%
\caption{\it Dependence of the branching ratio observables $\mathcal{B}(\kpn)$ (left) and $\mathcal{B}(\klpn)$ (right) on the CKM parameter inputs $|V_{cb}|$, $|V_{ub}|$ and $\gamma$. The 95\% C.L. bands in $V_{ub}$, $V_{cb}$ and $\gamma$ are shown in green, blue, and red, respectively.
\label{fig:CKMdependence}}
\end{figure}

For convenience we give the following parametric expressions for the branching ratios in terms of the CKM inputs:
\begin{align}
    \mathcal{B}(\kpn) = (8.39 \pm 0.30) \times 10^{-11} \cdot
    \bigg[\frac{\left|V_{cb}\right|}{40.7\times 10^{-3}}\bigg]^{2.8}
    \bigg[\frac{\gamma}{73.2^\circ}\bigg]^{0.74},\label{kplusApprox}
\end{align}
\begin{align}
    \mathcal{B}(\klpn) = (3.36 \pm 0.05) \times 10^{-11} \cdot
    &\bigg[\frac{\left|V_{ub}\right|}{3.88\times 10^{-3}}\bigg]^2
    \bigg[\frac{\left|V_{cb}\right|}{40.7\times 10^{-3}}\bigg]^2
    \bigg[\frac{\sin(\gamma)}{\sin(73.2^\circ)}\bigg]^{2}.
\end{align}
The parametric relation for $\mathcal{B}(\klpn)$ is exact, while for $\mathcal{B}(\kpn)$ it gives an excellent approximation:
for the large ranges $37 \leq |V_{cb}|\times 10^{3} \leq 45$ and $60^\circ \leq \gamma \leq 80^\circ$ it is accurate to 1\% and 0.5\%, respectively.
In the case of $\mathcal{B}(\kpn)$ we have absorbed $|V_{ub}|$ into the non-parametric error due to the weak dependence on it. 
The exact dependence of both branching ratios on $\vub$, $\vcb$ and $\gamma$ is 
shown in Figure~\ref{fig:CKMdependence}.

In order to obtain the values of $\varepsilon_K$, $S_{\psi K_S}$, $\Delta M_{s,d}$ and of the branching ratios for $B_{s,d}\to\mu^+\mu^-$ we use the known 
expressions collected in \cite{Buras:2013ooa}, together with the parameters listed in Table~\ref{tab:input}.
The ``bar'' on the $\Bsmumu$ branching ratio, $\overline{\cal B}(\Bsmumu)$, denotes an average over the two mass-eigenstates, as measured by experiment, rather than an average over the two flavour-states, which differs in the $B_s$ system\cite{DescotesGenon:2011pb,DeBruyn:2012wj,DeBruyn:2012wk}.

In Table~\ref{tab:SA} we show the results for the $\kpn$ and $\klpn$  branching ratios and other observables, for three choices of the pair $(\vub,\vcb)$ 
corresponding to the exclusive determination (\ref{exclusive}), the inclusive 
determination (\ref{inclusive}) and our average (\ref{average}). We use 
(\ref{gamma}) for $\gamma$ in each case. 
We observe:
\begin{itemize}
\item
The uncertainty in $\mathcal{B}(K^+ \to \pi^+\nu\bar\nu)$ amounts to more than 
$10\%$ and has to be decreased to compete with future NA62 measurements, but 
finding this branching ratio in the ballpark of $15\times 10^{-11}$ would clearly indicate NP at work. 
\item
On the other hand, consistency with $\overline{\mathcal{B}}(B_s \to \mu^+\mu^-)$ 
would imply the $\kpn$ branching ratio to be in the ballpark of $7\times 10^{-11}$. In such a case the search for NP in this decay will be a real challenge 
and the simultaneous measurement of $\klpn$ will be crucial.
\item
The values of $S_{\psi K_S}$ are typically above the data but only in the case 
of the {\it inclusive} determinations of both $\vcb$ and $\vub$ 
is a new CP phase required. 
\item
The accuracy on the SM prediction for $\Delta M_s$ and $\Delta M_d$ is
far from being satisfactory. Yet, the prospects of improving the accuracy 
by a factor of two to three in this decade are good.
\end{itemize}

\begin{table}[!tb]
\begin{center}
\renewcommand{\arraystretch}{1.5}\begin{tabular}{|c|ccc|c|}
\hline\hline
 & exclusive & inclusive & average & measured \\
\hline
${\rm BR}(K^+ \to \pi^+\nu\bar\nu)\ [10^{-11}]$ & ${7.62}^{+0.69}_{-0.70}$ & ${9.30}^{+0.89}_{-0.92}$ & ${8.39}^{+1.06}_{-1.03}$ & $17.3^{+11.5}_{-10.5}$\\
${\rm BR}(K_L \to \pi^0\nu\bar\nu)\ [10^{-11}]$ & ${2.88}^{+0.30}_{-0.35}$ & ${4.64}^{+0.63}_{-0.68}$ & ${3.36}^{+0.60}_{-0.61}$ & $\leq 2600$\\
$\overline{\rm BR}(B_s \to \mu^+\mu^-)\ [10^{-9}]$ & ${3.18}^{+0.18}_{-0.18}$ & ${3.66}^{+0.21}_{-0.20}$ & ${3.40}^{+0.28}_{-0.27}$ & $2.8 \pm 0.7$\\
${\rm BR}(B_d \to \mu^+\mu^-)\ [10^{-10}]$ & ${1.00}^{+0.11}_{-0.12}$ & ${1.17}^{+0.14}_{-0.14}$ & ${1.08}^{+0.13}_{-0.14}$ & $3.6^{+1.6}_{-1.4}$\\
$|\varepsilon_K|\ [10^{-3}]$ & ${1.96}^{+0.25}_{-0.27}$ & ${2.74}^{+0.36}_{-0.38}$ & ${2.23}^{+0.35}_{-0.36}$ & $2.228\pm 0.011$\\
$S_{\psi K_S}^{\rm SM}$ & ${0.74}^{+0.02}_{-0.03}$ & ${0.80}^{+0.03}_{-0.04}$ & ${0.75}^{+0.05}_{-0.05}$ & $0.682 \pm 0.019$\\
$\Delta M_s\ [{\rm ps}^{-1}]$ & ${16.19}^{+2.37}_{-2.23}$ & ${18.64}^{+2.73}_{-2.56}$ & ${17.34}^{+2.74}_{-2.58}$ & $17.761 \pm 0.022$\\
$\Delta M_d\ [{\rm ps}^{-1}]$ & ${0.52}^{+0.09}_{-0.09}$ & ${0.60}^{+0.11}_{-0.11}$ & ${0.55}^{+0.10}_{-0.10}$ & $0.510 \pm 0.003$\\
${\rm Im}(\lambda_t)\ [10^{-4}]$ & ${1.40}^{+0.07}_{-0.09}$ & ${1.78}^{+0.12}_{-0.13}$ & ${1.51}^{+0.13}_{-0.14}$ & $-$\\
${\rm Re}(\lambda_t)\ [10^{-4}]$ & ${-2.99}^{+0.19}_{-0.19}$ & ${-3.39}^{+0.24}_{-0.23}$ & ${-3.20}^{+0.29}_{-0.29}$ & $-$\\
$R_b$ & ${0.41}^{+0.02}_{-0.02}$ & ${0.45}^{+0.03}_{-0.03}$ & ${0.41}^{+0.03}_{-0.03}$ & $-$\\
\hline\hline
\end{tabular}
\end{center}
\caption{\it Values of $\mathcal{B}(\kpn)$, $\mathcal{B}(\klpn)$ and 
of other observables within the SM for the three choices of $\vub$ and $\vcb$
following strategy A as discussed in the text.
\label{tab:SA}}
\end{table}


\subsection{Correlations between observables}

\boldmath
\subsubsection*{Correlations between $\kpn$ and $\Bqmumu$}
\unboldmath

From inspection of the formulae for the branching ratios for $\kpn$ and 
$B_{s,d}\to\mu^+\mu^-$ , each of which in particular depends on $\vcb$, we derive the following approximate relations 
\begin{align}
\mathcal{B}(\kpn) &= (8.39\pm 0.58)\times 10^{-11} \cdot  \left[\frac{\gamma}{73.2^\circ}\right]^{0.81}\notag\\
&\qquad\qquad\qquad\quad\times\left[\frac{\overline{\mathcal{B}}(B_s\to\mu^+\mu^-)}{3.4\times 10^{-9}}\right]^{1.42}\left[\frac{227.7\mev}{F_{B_s}}\right]^{2.84},\label{master1}\\
\mathcal{B}(\kpn) &= (8.41 \pm 0.77)\times 10^{-11} \cdot \left[\frac{\overline{\mathcal{B}}(B_s\to\mu^+\mu^-)}{3.4\times 10^{-9}}\right]^{0.74}\left[\frac{227.7\mev}{F_{B_s}}\right]^{1.48}\notag\\
&\qquad\qquad\qquad\quad\times\left[\frac{\mathcal{B}(B_d\to\mu^+\mu^-)}{1.08\times 10^{-10}}\right]^{0.72}\left[\frac{190.5\mev}{F_{B_d}}\right]^{1.44}.\label{master2}
\end{align}
Note that both relations are independent of $\vcb$ and (\ref{master1}) depends only on $\gamma$.
In particular the correlation (\ref{master1}) should be of interest in 
the coming years due to  
the measurement of $\kpn$ by NA62, of $B_s\to\mu^+\mu^⁻$ by LHCb and CMS and of $\gamma$ by LHCb. Moreover the last factor should also be improved by lattice 
QCD.

\begin{figure}[t]
\centering%
\includegraphics[width=0.455\textwidth]{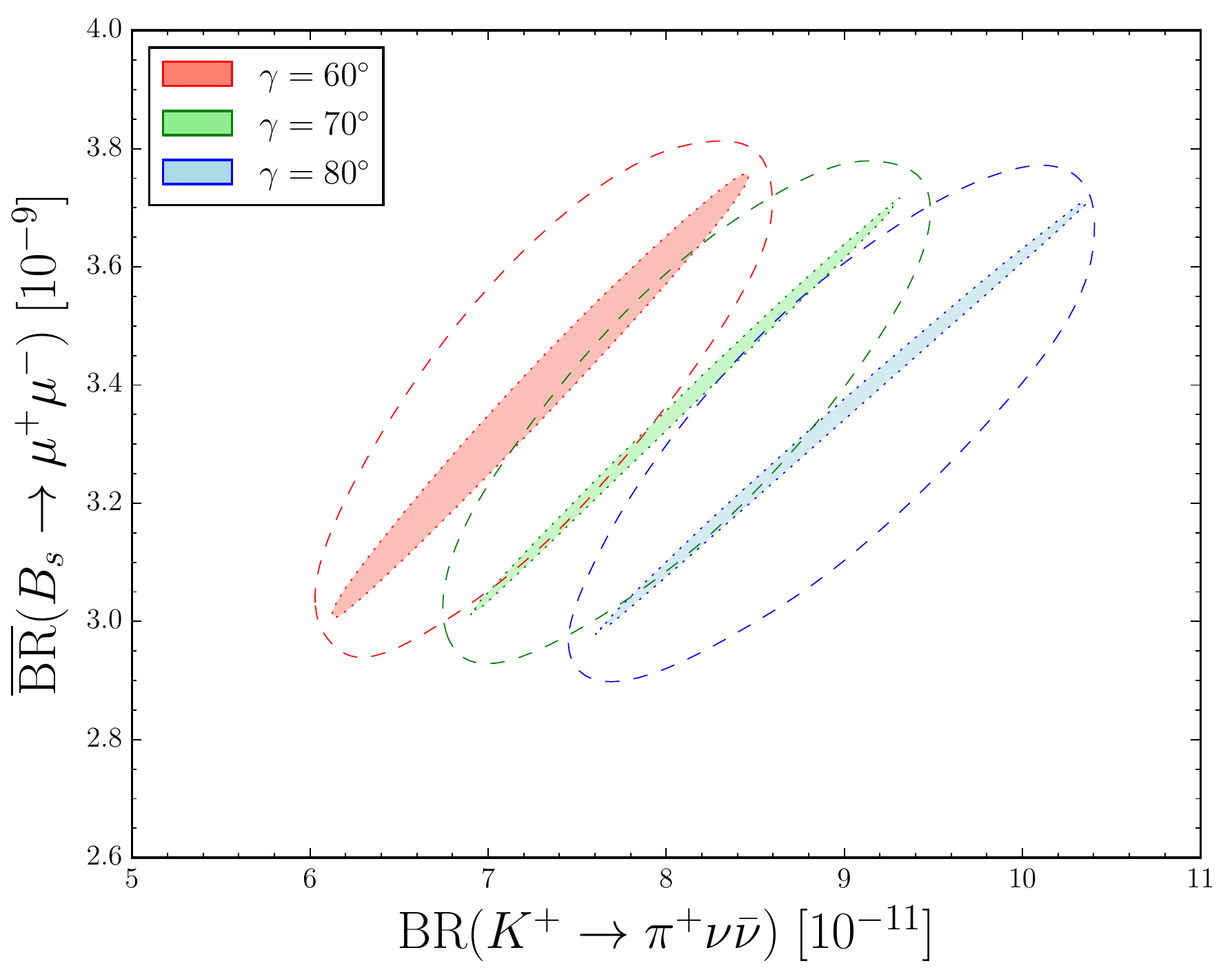}
\includegraphics[width=0.45\textwidth]{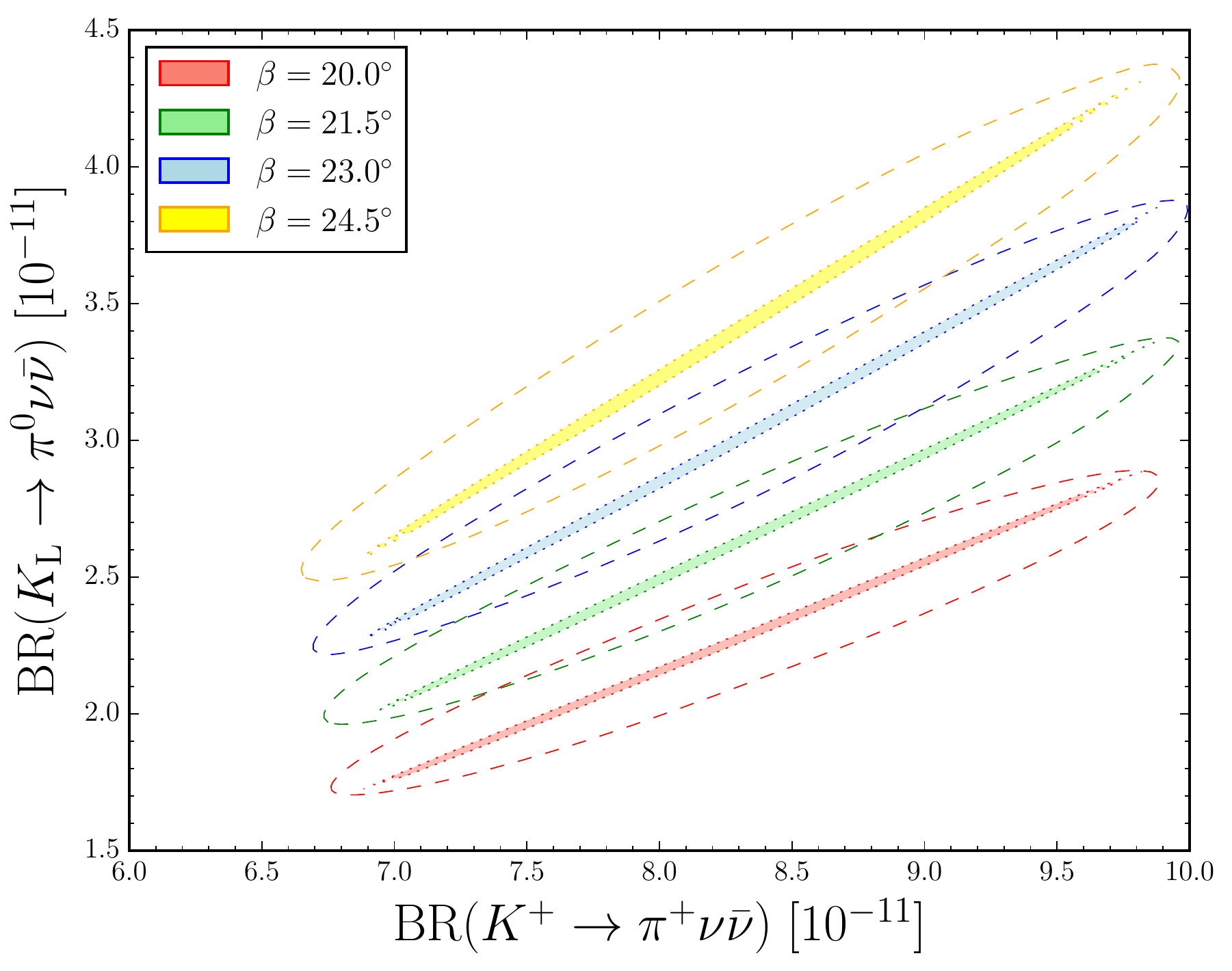}%
\caption{\it Left panel: correlation of $\overline{\mathcal{B}}(B_s\to\mu^+\mu^-)$ versus $\mathcal{B}(\kpn)$ for fixed values of $\gamma$. 
Right panel: correlation of ${\mathcal{B}}(\klpn)$ versus $\mathcal{B}(\kpn)$ for fixed values of $\beta$. 
In both plots the dashed regions correspond to a 68\% C.L. resulting from the uncertainties on all other inputs, 
while the inner filled regions result from including only the uncertainties from the remaining CKM inputs of strategy A.\label{fig:fixedPlots}}
\end{figure}

In the left panel of Figure~\ref{fig:fixedPlots} we show the correlation between $\kpn$ and $B_s\to\mu^+\mu⁻$
for different fixed values of $\gamma$.
The dashed regions correspond to a 68\% C.L. that results from including the uncertainties on all the other input parameters, whereas the inner filled regions are a result of only including the uncertainties of $\vub$, $\vcb$ (we use the averages in \eqref{average}), and $\vus$.

It should be noticed that the present experimental determination of $\overline{\mathcal{B}}(B_s\to\mu^+\mu^-)$ is slightly lower than the SM prediction, and the agreement between the SM and the data can be improved by lowering $\vcb$ to values in the ballpark of its present exclusive determinations. But in this case, as can be seen already from Table~\ref{tab:SA}, the SM predictions for both $\B(\kpn)$ and $\varepsilon_K$ are also reduced. It can be useful to express $\B(\kpn)$ as a function of $\varepsilon_K$, in a way similar to \eqref{master1} and \eqref{master2}, in order to make the correlation between them explicit. One has
\begin{align}
\B(\kpn) &= (8.39\pm 1.11)\times 10^{-11}\cdot \left[\frac{|\varepsilon_K|}{2.23\times 10^{-3}}\right]^{1.07}\notag\\
&\qquad\qquad\qquad\quad \times\left[\frac{\gamma}{73.2^\circ}\right]^{-0.11}\cdot\left[\frac{|V_{ub}|}{3.88\times 10^{-3}}\right]^{-0.95}.\label{master3}
\end{align}
We do not write explicitly the dependence on the hadronic quantities, since here more parameters are involved. The uncertainty here comes mainly from $\eta_{cc}$ and $\eta_{ct}$, while the ones due to $F_K$ are smaller than the corresponding ones in the $B_{s,d}$ meson systems.
It is evident from this formula that a reduction of $\B(\kpn)$ implies also a reduction of $\varepsilon_K$.

The correlations in (\ref{master1}), (\ref{master2}) and (\ref{master3}) result from 
the fact that it is possible, by taking suitable powers of the $B_{s,d}\to \mu^+\mu^-$ branching ratios, to eliminate the dependence on $\vcb$, while the one-loop functions $X$, $Y$, and $S$ are fixed by the top mass in the SM.
Both correlations could be broken already in models with constrained 
MFV (CMFV) in which the modifications of the functions $X$ and $Y$ are generally different. In general MFV models new 
scalar operators could additionally contribute to $B_{s,d}\to \mu^+\mu^-$, modifying also
the factors involving the weak decay constants. Therefore, these correlations are strictly valid only in the SM, and their violation would not necessarily rule out (C)MFV.

\boldmath
\subsubsection*{$\kpn$ and $\klpn$ in Minimal Flavour Violation}
\unboldmath

In models of NP with Minimal Flavour Violation (MFV) there are no flavour-changing interactions beyond those generated from the SM Yukawa couplings \cite{D'Ambrosio:2002ex}.
For $\kpn$ and $\klpn$ this restricts the operators that can contribute in most NP models to just the operator already dominant in the SM, $(\bar s d)_{V-A}(\bar\nu\nu)_{V-A}$, making MFV equivalent to Constrained MFV (CMFV) \cite{Buras:2000dm} in this case.
Therefore in MFV the value of $X(x_t)$ can be shifted but must stay real.
Defining for convenience
\begin{align}\label{b1b2}
    B_+&=\frac{\mathcal{B}(\kpn)}{\kappa_+(1+\Delta_{\rm EM})},&
B_L&=\frac{\mathcal{B}(\klpn)}{\kappa_L},
\end{align}
with $\kappa_+$ and $\kappa_L$ given in (\ref{kapp}) and (\ref{kapl}), respectively,
we have in the case of MFV the correlation
\begin{equation}
    B_+ = B_{\rm L} + \left[\frac{\RE\lambda_t}{\IM\lambda_t}\sqrt{B_{\rm L}} - \left(1-\frac{\lambda^2}{2}\right){\rm sgn}\left(X(x_t)\right) P_c(X) \right]^2, \label{genMFVrel}
\end{equation}
which was first given in \cite{Buchalla:1994tr,Buras:2001af}. Note that in the SM the sign of $X(x_t)$ is positive.
Recalling the relation
\begin{equation}\label{betaRel}
    \frac{\RE\lambda_t}{\IM\lambda_t}  \simeq {-\cot\beta}\left(1-\frac{\lambda^2}{2}\right)^2,
\end{equation}
accurate to $\lambda^4$ terms,
and solving for $\beta$ then gives
\begin{equation}\label{cbb}
\cot\beta = \frac{1}{\left(1-\frac{\lambda^2}{2}\right)^{2}}\left[\sqrt{\frac{B_+-B_L}{B_L}}-\left(1 - \frac{\lambda^2}{2}\right)\frac{{\rm sgn}\left(X(x_t)\right) P_c(X)}{\sqrt{B_L}}\right].
\end{equation}

In deriving (\ref{cbb}) from (\ref{genMFVrel}) and (\ref{betaRel}) one finds that for $X(x_t)>0$ this solution is unique, while for $X(x_t)<0$ a second solution with a minus sign in front of first square root is allowed \cite{Buras:2001af}. However this solution is excluded if we require both branching ratios to be larger than $10^{-11}$ and we will not consider it here.

In the SM and CMFV we have to a very good approximation the relation~\cite{Buchalla:1994tr,Buras:2001af} 
\begin{equation}\label{sin}
S_{\psi K_S}=\sin 2\beta,
\end{equation}
which is only spoiled by possible penguin enhancements in the $B_d\to J/\psi K_{\rm S}$ mode~\cite{Faller:2008zc}.
Thus \eqref{sin} together with \eqref{cbb} give a triple correlation between $\kpn$, $\klpn$ and $S_{\psi K_S}$ in the SM and CMFV.

As demonstrated in the earlier parts of this section, the branching ratios for $\kpn$ and $\klpn$ still contain significant parametric uncertainties due to the uncertainties in 
$\vcb$, $\vub$ and $\gamma$, and to a lesser extent in $m_t$. 
It is therefore remarkable that within the SM all these uncertainties practically cancel out in this triple correlation~\cite{Buchalla:1994tr}. 
Moreover, this property turns out to be true for all  models with constrained MFV
\cite{Buras:2001af}.

We note that the main uncertainty in (\ref{cbb}) resides in $P_c$, as the 
uncertainty in $\lambda$ is very small. We stress that this relation is
practically immune to any variation of the function $X(x_t)$ within MFV 
models. This means that once $\mathcal{B}(\kpn)$ and $S_{\psi K_S}$ will 
be precisely measured we will know the {\it unique} value of $\mathcal{B}(\klpn)$ 
within CMFV models. 
This relation is analogous to the one between $\mathcal{B}(B_{s,d}\to\mu^+\mu^-)$ 
and $\Delta M_{s,d}$ \cite{Buras:2003td}, where the present 
knowledge of  $\Delta M_{s,d}$ together with the future precise value of 
$\mathcal{B}(B_s\to\mu^+\mu^-)$ will allow us to uniquely predict the 
branching ratio $\mathcal{B}(B_d\to\mu^+\mu^-)$ in CMFV models well ahead 
of its precise direct measurement.

In the right panel of Figure~\ref{fig:fixedPlots} we show the correlation between $\kpn$ and $\klpn$ for different fixed values of $\beta$ ($S_{\psi K_S}$).
The dashed regions correspond to a 68\% C.L. that results from including the uncertainties on all the other input parameters, whereas the inner filled regions are a result of only including the uncertainties of $\vcb$ (we use the average in \eqref{average}), $\gamma$ (as given in \eqref{gamma}) and $\vus$ in \eqref{recentInputs}.
We observe that in the latter case the dependence on the remaining CKM parameters, for fixed $\beta$, is indeed minimal.

\begin{figure}[t]
\centering%
\includegraphics[width=0.45\textwidth]{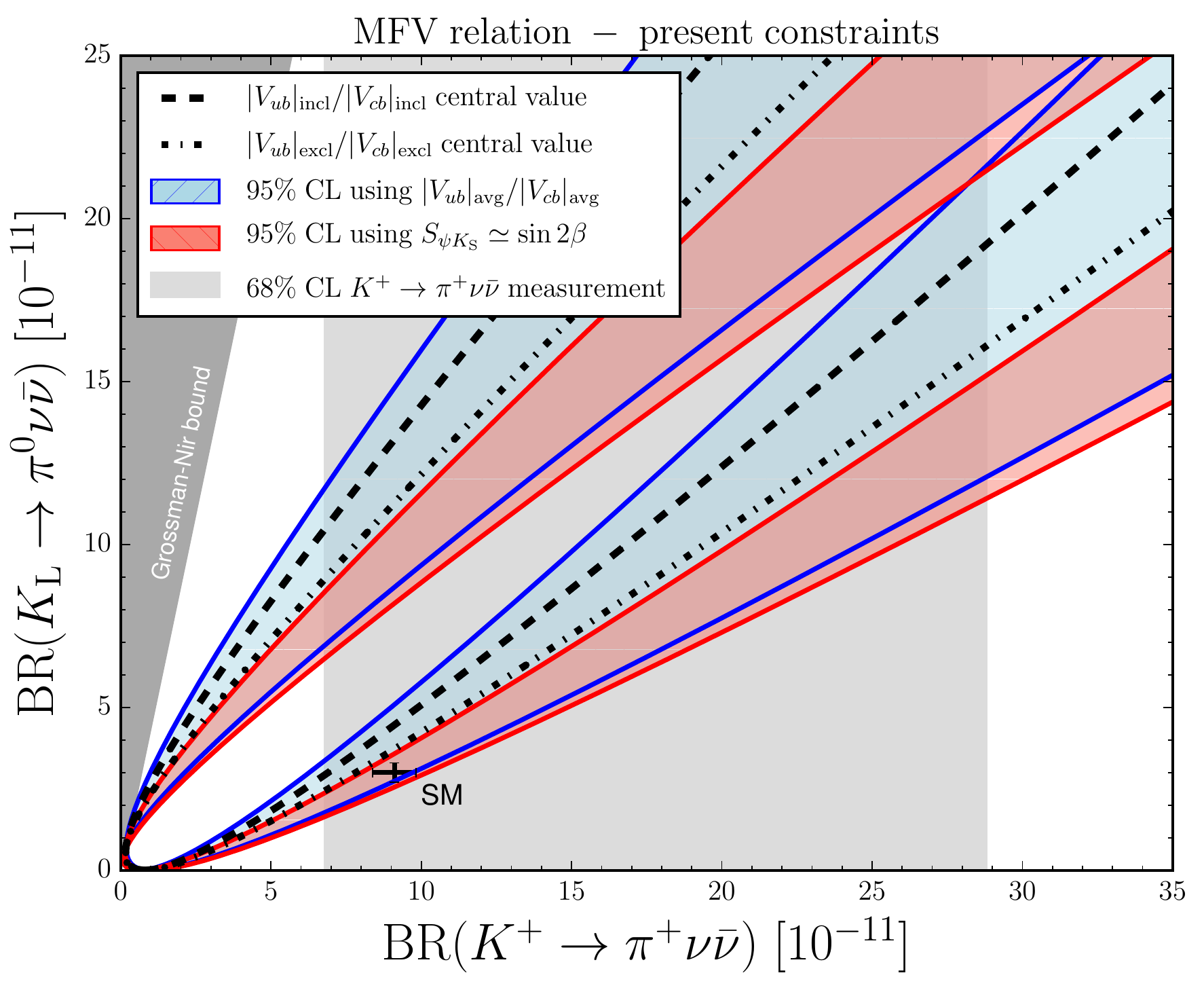}%
\includegraphics[width=0.45\textwidth]{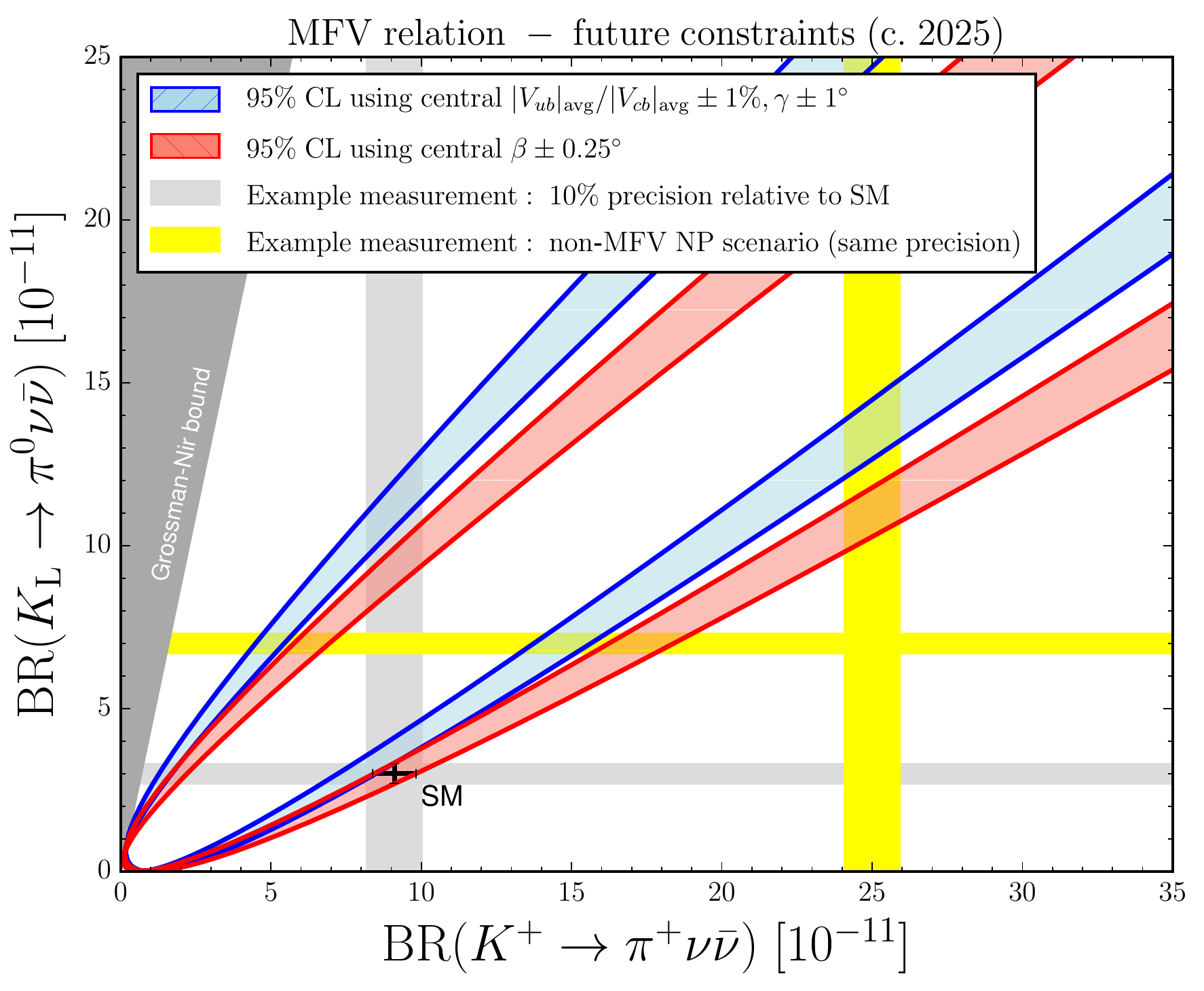}%
\caption{\it The MFV relation between $\kpn$ and $\klpn$ using $S_{\psi K_{\rm S}}\simeq \sin 2\beta$ versus using the various tree-level inputs of $|{V_{cb}}/{V_{ub}}|$ and $\gamma$ (see text). In the left panel we show situation from current constraints, and in the right panel the possible situation in the following decade, including 10\% precision on the two branching ratios, for illustration.\label{fig:RbMFV}}
\end{figure}

It is also possible to express the ratio in (\ref{betaRel}) as
\begin{equation}\label{treeMFVrel}
    \frac{\RE\lambda_t}{\IM\lambda_t} = \frac{\left(1 - \frac{\lambda^2}{2}\right)^2}{\sin\gamma} \left[\cos\gamma - \left(\frac{\lambda}{1 - \frac{\lambda^2}{2}}\right)\left|\frac{V_{cb}}{V_{ub}}\right| \right],
\end{equation}
i.e.\ in terms of the tree-level CKM inputs discussed in this section, which are generally assumed to be free of NP effects.
We note that in MFV also $S_{\psi K_S}$ is not affected by NP and is more accurately determined than $\gamma$.
On the other hand, there is a class of models -- e.g.\ models with a $U(2)^3$ flavour symmetry \cite{Barbieri:2012uh} -- where the correlation with $S_{\psi K_S}$ is no longer true, while \eqref{genMFVrel} and the generic relation \eqref{treeMFVrel} still hold.

While the virtue of the correlation (\ref{cbb}) is its very weak dependence on the CKM parameters, the correlation  (\ref{genMFVrel}) together with (\ref{treeMFVrel}) shares partly this property as it depends only on the ratio $|{V_{cb}}/{V_{ub}}|$, equivalent to $R_b$, and not on $\vub$ and $\vcb$ separately.
As we can see from the values of $R_b$ given in Table~\ref{tab:SA}, this avoids some of the trouble with exclusive versus inclusive determinations, as the ratio of purely exclusive or inclusive determinations, as well as their weighted average, results in less variation -- i.e. only 5\% among the cases considered.
Note that combining exclusive $|V_{ub}|$ with inclusive $|V_{cb}|$, for example, gives a greater variation.

In the left panel of Figure~\ref{fig:RbMFV} we compare the MFV relation for various values of $|{V_{cb}}/{V_{ub}}|$, including a $1\,\sigma$~C.L. region corresponding to our weighted averages.
We also include for comparison the relation corresponding to the current $S_{\psi K_{\rm S}}$ measurement, which is seen to be more accurate.
In the right panel we repeat this comparison for possible future constraints in the following decade, where our choice of errors are based on those collected in~\cite{Buras:2014zga}.
For the branching ratios of $\kpn$ and $\klpn$ we assume a 10\% precision relative to the SM predictions.
Though this is the realistic target set by the NA62 experiment for $\kpn$, the KOTO experiment will likely not reach such a precision for $\klpn$. 
We observe that in this possible sketch of the future, the two decays under consideration have the potential to probe  MFV and/or a $U(2)^3$ symmetry.


\section{CKM inputs from loop-level observables}\label{sec:B}

A different approach is to assume that there are no relevant NP contributions to all the quantities 
listed in (\ref{STRB}), so that we can use them together with the precise 
value of $\vus$ to determine the best values of 
\be\label{OUTPUT}
\beta,\qquad \vcb, \qquad \vub, \qquad \vtd, \qquad \vts,
\ee
and predict the branching ratios for $\kpn$, $\klpn$ and $B_{s,d}\to\mu^+\mu^-$. Clearly the absence of NP effects in all the loop observables \eqref{STRB} requires the SM to be valid up to a reasonably high energy scale, which is a stronger assumption with respect to the one of strategy A, where only tree-level 
determinations of CKM parameters  were assumed to be free of NP effects.
We call this approach strategy~B.

The relevant SM expressions can be found in \cite{Buras:2013ooa} and in 
particular in  \cite{Buras:2013raa}, where this strategy has been used to 
determine the correlation between the values of $\vcb$ and $\vub$ with the 
non-perturbative parameters relevant for $\Delta M_{s,d}$. As the precision 
on these parameters resulting from QCD lattice calculations is improving,
and the value of $\hat B_K$, relevant for $\varepsilon_K$, has been known precisely already for some time,\footnote{We use $\hat B_K=0.750(15)$ which takes into account the values obtained by lattice QCD \cite{Aoki:2013ldr} and 
large N approach \cite{Buras:2014maa}.} we can now use these formulae to extract the values 
listed in (\ref{OUTPUT}). 

At least three independent observables among the four listed in \eqref{STRB} have to be used in order to fix the three free parameters of the CKM matrix besides $|V_{us}|$. For illustration, we present here the strategy which allows us to determine the parameters in \eqref{OUTPUT} with high precision with the minimal number of measurements. Schematically this procedure can be described in two steps:

\begin{itemize}
 \item {\bf Step 1:}
The unitarity triangle can be determined from the experimental values of $S_{\psi K_S}=\sin 2\beta$ and the mass ratio
\begin{align}\label{DMqRatio}
    \frac{\Delta M_d}{\Delta M_s}
    &= \frac{m_{B_d}}{m_{B_s}}\frac{1}{\xi^2} \frac{\vtd^2}{\vts^2}, &
    \xi&\equiv\frac{F_{B_s}\sqrt{\hat B_{B_s}}}{F_{B_d}\sqrt{\hat B_{B_d}}}.
\end{align}
Using the following very accurate expressions for $|V_{td}|$ and $|V_{ts}|$,
\begin{align}\label{RT}
|V_{td}| &\simeq \lambda |V_{cb}| R_t, & |V_{ts}| &\simeq \left(1 + \frac{\lambda^2}{2}(1 - 2R_t\cos\beta)\right)|V_{cb}|,
\end{align}
where $R_t$ is one of the sides of the UT, and solving \eqref{DMqRatio} for $R_t$ one gets
\begin{equation}\label{vtqExpr}
R_t \simeq \frac{\xi}{|V_{us}|}\sqrt{\frac{\Delta M_d}{\Delta M_s}\frac{m_{B_s}}{m_{B_d}}}\left(1 - |V_{us}|\xi\sqrt{\frac{\Delta M_d}{\Delta M_s}\frac{m_{B_s}}{m_{B_d}}} + \frac{|V_{us}|^2}{2} + \cdots\right)
\end{equation}
where the dots indicate terms of order $\mathcal{O}(|V_{us}|^4,|V_{us}|^2\Delta M_d/\Delta M_s)$.

With one side, $R_t$, and one angle, $\beta$, known, the full unitarity triangle is determined by means of purely geometrical relations. In particular one has
\begin{align}\label{VUBG}
R_b &= \frac{1-\lambda^2/2}{\lambda}\left|\frac{V_{ub}}{V_{cb}}\right| = \sqrt{1+R_t^2-2 R_t\cos\beta},&
\cot\gamma&=\frac{1-R_t\cos\beta}{R_t\sin\beta},
\end{align}
and the apex $(\bar\varrho,\bar\eta)$ 
of the triangle is given by 
\be\label{UT}
\bar\varrho=1-R_t\cos\beta, \qquad \bar\eta=R_t\sin\beta\, .
\ee
The very precise value of $R_t$ obtained through $\Delta M_d/\Delta M_s$ therefore allows a very precise determination of $\gamma$.

It should be emphasised that the UT constructed in this manner is universal for all CMFV models as the box function $S$ does not enter the expressions used in these two steps \cite{Buras:2000dm}. Moreover this determination is independent of $\vcb$.

\item {\bf Step 2:} 
The measured value of $|\varepsilon_K|$ then allows us to determine the optimal value of $\vcb$. 
Indeed we have \cite{Buras:2008nn}
\begin{align}\label{epsilonK}
|\varepsilon_K^{\rm SM}| =\, &\kappa_\varepsilon \frac{G_F^2 F_K^2 m_{K^0} M_W^2}{6 \sqrt{2} \pi^2 \Delta M_K} \hat B_K\vcb^2 \lambda^2 R_t \sin\beta,\notag\\
&\times\left( \vcb^2 R_t \cos\beta\, \eta_{tt} S_0(x_t) + \eta_{ct} S_0(x_c,x_t) - \eta_{cc} x_c \right)
\end{align}
where $x_i=m^2_i/M_W^2$, $S_0$ is the well known SM box function as defined {\it e.g.} in \cite{Buras:2013ooa},
and $\kappa_\varepsilon=0.94\pm0.02$ \cite{Buras:2008nn,Buras:2010pza}.
With $R_t$ known from \eqref{RT} and $\beta$ determined from $S_{\psi K_S}$, the only unknown in (\ref{epsilonK}) is $\vcb$. 
Having found $\vcb$, $R_b$, and $R_t$, also $\vub$, $\vtd$ and $\vts$ are determined by the previous relations in a straightforward way.
\end{itemize}

Alternatively we can also determine $\vcb$ by using separately $\Delta M_{s}\propto \vts^2$ and 
$\Delta M_{d} \propto \vtd^2$ instead of $\epsilon_K$, as both are proportional to $\vcb^2$ via the expressions given in \eqref{vtqExpr}.
In principle it is also possible to determine the CKM matrix from $\Delta M_d$, $\Delta M_s$, and $\epsilon_K$, but the precision in this case will be rather limited, due to the absence of the strong constraint on $\beta$ from $S_{\psi K_S}$.
The best accuracy is obtained by performing a simultaneous fit to all the four observables ${\Delta M_d}$, ${\Delta M_s}$, $S_{\psi K_{\rm S}}$ and $\epsilon_K$.

\begin{table}[!tb]
\begin{center}
\renewcommand{\arraystretch}{1.5}\scalebox{0.8}{
\begin{tabular}{|c|ccc|}
\hline\hline
 & $\{|\varepsilon_K|,\Delta M_d/\Delta M_s,S_{\psi K_{\rm S}}\}_{\rm SM}$ & $\{\Delta M_d,\Delta M_s,S_{\psi K_{\rm S}}\}_{\rm SM}$ & $\{|\varepsilon_K|,\Delta M_d,\Delta M_s,S_{\psi K_{\rm S}}\}_{\rm SM}$ \\
\hline
$|V_{cb}|\ [10^{-3}]$ & ${42.59}^{+1.41}_{-1.26}$ & ${41.30}^{+2.65}_{-2.47}$ & ${42.35}^{+1.25}_{-1.13}$\\
$|V_{ub}|\ [10^{-3}]$ & ${3.62}^{+0.15}_{-0.14}$ & ${3.51}^{+0.27}_{-0.25}$ & ${3.61}^{+0.15}_{-0.14}$\\
$|V_{td}|\ [10^{-3}]$ & ${8.96}^{+0.28}_{-0.28}$ & ${8.68}^{+0.66}_{-0.62}$ & ${8.95}^{+0.27}_{-0.28}$\\
$|V_{ts}|\ [10^{-3}]$ & ${41.79}^{+1.43}_{-1.27}$ & ${40.52}^{+2.60}_{-2.42}$ & ${41.55}^{+1.27}_{-1.14}$\\
${\cal B}(K^+ \to \pi^+\nu\bar\nu)\ [10^{-11}]$ & ${9.18}^{+0.79}_{-0.71}$ & ${8.39}^{+1.76}_{-1.41}$ & ${9.08}^{+0.74}_{-0.68}$\\
${\cal B}(K_L \to \pi^0\nu\bar\nu)\ [10^{-11}]$ & ${3.01}^{+0.33}_{-0.29}$ & ${2.66}^{+0.84}_{-0.63}$ & ${2.98}^{+0.32}_{-0.28}$\\
$\overline{\cal B}(B_s \to \mu^+\mu^-)\ [10^{-9}]$ & ${3.69}^{+0.30}_{-0.26}$ & ${3.46}^{+0.49}_{-0.43}$ & ${3.64}^{+0.27}_{-0.24}$\\
${\cal B}(B_d \to \mu^+\mu^-)\ [10^{-10}]$ & ${1.09}^{+0.08}_{-0.08}$ & ${1.02}^{+0.17}_{-0.15}$ & ${1.09}^{+0.08}_{-0.08}$\\
${\rm Im}(\lambda_t)\ [10^{-4}]$ & ${1.43}^{+0.08}_{-0.07}$ & ${1.35}^{+0.20}_{-0.17}$ & ${1.42}^{+0.07}_{-0.07}$\\
${\rm Re}(\lambda_t)\ [10^{-4}]$ & ${-3.46}^{+0.18}_{-0.19}$ & ${-3.25}^{+0.40}_{-0.45}$ & ${-3.43}^{+0.17}_{-0.18}$\\
\hline\hline
\end{tabular}
}
\end{center}
\caption{\it Results of the fit to the CKM matrix elements for various combinations of inputs as detailed in strategy B, and the corresponding observable predictions.\label{tab:SB}}
\end{table}

In Table~\ref{tab:SB} we give the results of fits for the CKM matrix elements using different combinations of the inputs discussed in the steps above. The values for the experimental and lattice observables used as inputs are listed in Table~\ref{tab:input}.
The fits were performed using a Bayesian statistical approach: uncorrelated Gaussian priors were chosen for each of the input parameters and the posterior distribution was sampled using Markov Chain Monte Carlo with the help of the Bayesian Analysis Toolkit~\cite{Caldwell:2008fw}. A direct minimisation of the $\chi^2$, yielding identical results, has also been performed as a check.
We observe that using $|\varepsilon_K|$ in step 2 gives a more precise result for $\vcb$ than the alternative of using $\Delta M_d$ and $\Delta M_d$ separately,
as well as favouring a higher central value.
The most accurate determination (given in the last column of the table), follows from including all inputs.
The corresponding CKM matrix elements of interest are:
\begin{align}
    |V_{ub}| &= (3.61 \pm 0.14) \times 10^{-3}, &
    |V_{cb}| &= (42.4 \pm 1.2) \times 10^{-3},\notag\\
    |V_{td}| &= (8.94 \pm 0.27) \times 10^{-3}, &
    |V_{ts}| &= (41.6 \pm 1.2) \times 10^{-3}. 
\end{align}

For completeness, we give here the sides of the UT as determined from our full fit, that read
\begin{align}
R_t &= 0.937 \pm 0.032, & R_b &= 0.368 \pm 0.013,
\end{align}
while its angles are
\begin{align}
\alpha &= (89.0 \pm 5.0)^\circ, & \beta &= (21.5 \pm 0.8)^\circ, & \gamma &= (69.5 \pm 5.0)^\circ,
\end{align}
and its apex
\begin{align}
\bar\varrho &= 0.129 \pm 0.030, & \bar\eta &= 0.344 \pm 0.017.
\end{align}

The precision on $R_t$, $\gamma$ and $|V_{cb}|$ using the above strategy is already impressive, and will continue to improve with new lattice results.
Using for instance the improved error estimates for $\xi$ and $f_{B_s} \sqrt{\hat{B}_{B_s}}$ from \cite{Bouchard:2014eea} (keeping the central values from \cite{Aoki:2013ldr}) we find the very precise results:
\begin{align}\label{newLatticeErrRes}
    |V_{cb}| &= (42.0 \pm 0.9)\times 10^{-3},&
    \gamma &= (70.8 \pm 2.3)^\circ, &
    R_t &= 0.945 \pm 0.015.
\end{align}

\begin{figure}[t]
\centering%
\includegraphics[width=0.6\textwidth]{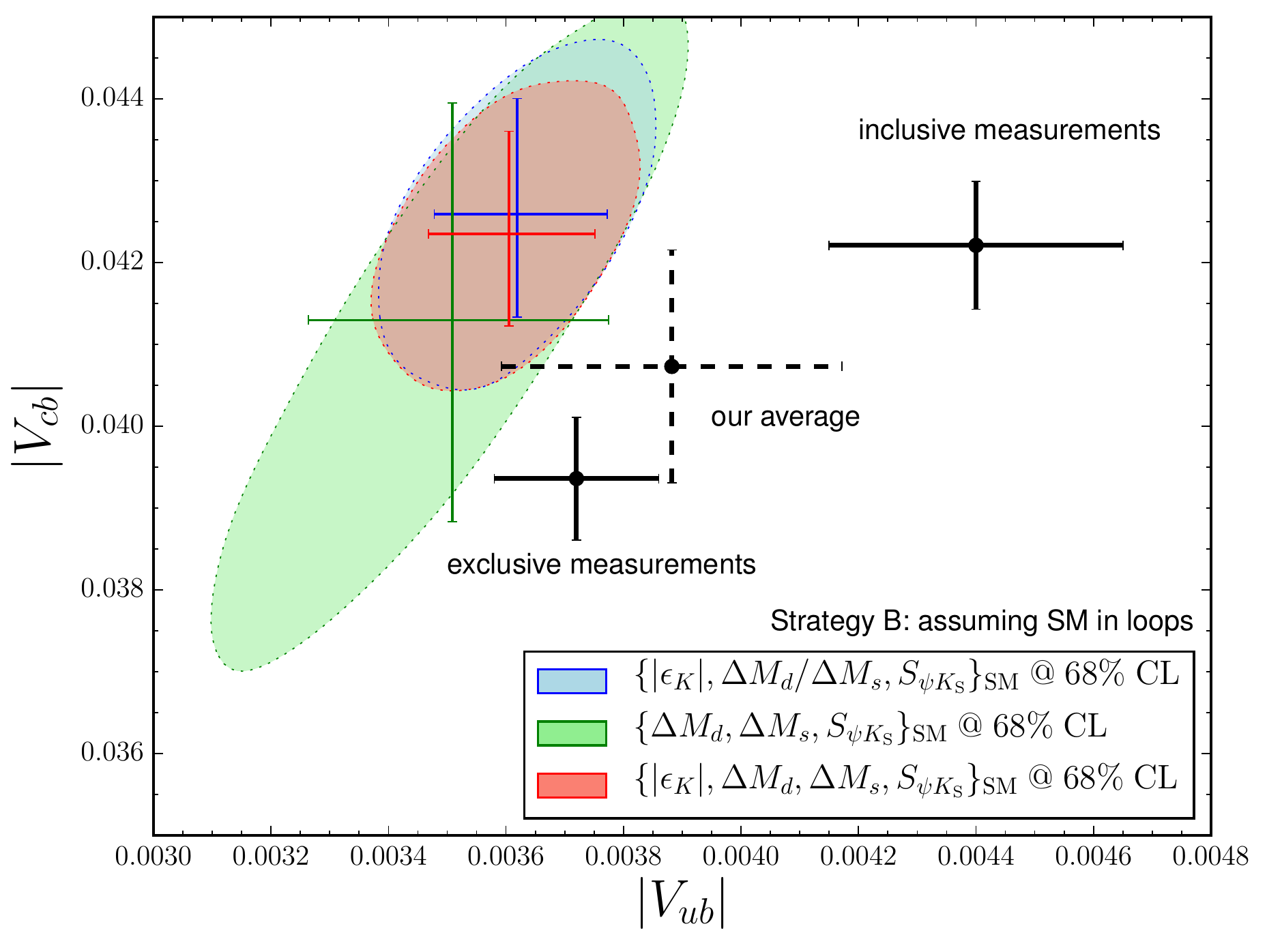}%
\caption{\it Comparison of 68\% C.L. regions for $\vub$ and $\vcb$ in strategy B for various combinations of inputs versus their reported inclusive and exclusive values, and our averages of these, as considered in strategy A.\label{fig:grandVxbPlots}}
\end{figure}

In Figure~\ref{fig:grandVxbPlots} we show the fitted ranges for $\vub$ and $\vcb$ and compare them with the inclusive, exclusive and our averaged
values in (\ref{exclusive})--(\ref{average}).
We distinguish between three different cases: the blue area corresponds to the fitted range of $\vub$ and $\vcb$ determined by $|\varepsilon_K|$, $\Delta M_d/\Delta M_s$ and $S_{\psi K_S}$; for the green area $\Delta M_d,\,\Delta M_s$ and $S_{\psi K_S}$ are used as inputs and 
the red area combines both and uses $|\varepsilon_K|$, $\Delta M_d,\,\Delta M_s$ and $S_{\psi K_S}$ as inputs.
As noted earlier, one can see that especially $|\varepsilon_K|$ favours large values of $\vcb$, around the inclusive value,  while the rather small $\vub$, around the exclusive value, is favoured by $S_{\psi K_S}$.

It is interesting to compare these results with the indirect fits performed by UTfit \cite{Bona:2006ah} and CKMfitter \cite{Charles:2015gya}, which give
\begin{align}
 \text{UTfit: }\quad\vub &=(3.63\pm0.12)\times 10^{-3}, &  \vcb&=(41.7\pm0.56)\times 10^{-3}\,,\label{equ:UTfit}\\
\text{CKMfitter: }\quad\vub &=\left(3.55^{+0.17}_{-0.15}\right)\times 10^{-3}, &  \vcb&=\left(41.17^{+0.90}_{-1.14}\right)\times 10^{-3}\,.\label{equ:CKMfitter}
\end{align}
They are in very good agreement with our results. We note however, that these two groups 
included in their analyses the information from tree level decays,  which we have decided not to include in our strategy B because of the discrepancies between inclusive and 
exclusive determinations of $\vub$ and $\vcb$. Moreover, we also did not use 
the tree-level determination of $\gamma$  contrary to these two groups.

Having determined the full CKM matrix in this manner, predictions for rare decays branching ratios can be made. These are collected in the last four rows in Table~\ref{tab:SB} and again the most precise are the ones in the last column 
so that our final results for the four branching ratios  are:
\begin{align}
\mathcal{B}(\kpn) &= \left(9.11\pm 0.72\right) \times 10^{-11},\label{kpnnfinal}\\
\mathcal{B}(\klpn) &= \left(3.00 \pm 0.31\right) \times 10^{-11},\\
\overline{\mathcal{B}}(B_s\to\mu^+\mu^-) &= \left(3.66\pm 0.26\right) \times 10^{-9},\label{bsmumufinal}\\
\mathcal{B}(B_d\to\mu^+\mu^-) &= \left(1.09\pm 0.08 \right) \times 10^{-10}\label{bdmumufinal}.
\end{align}
In \eqref{newLatticeErrRes} we used the new lattice error estimates from \cite{Bouchard:2014eea} for a ``sneak preview'' of how the CKM fit in strategy B will improve once the full results will be available.
Using these results for the observable predictions listed above will likewise lead to reduced uncertainties:  $\delta\mathcal{B}(\kpn) = 0.65$, $\delta\mathcal{B}(\klpn) = 0.28$, $\delta\mathcal{B}(B_s\to\mu^+\mu^-) = 0.22$ and $\delta\mathcal{B}(B_d\to\mu^+\mu^-) = 0.07$.

As a comparison, using instead the fit results of \eqref{equ:UTfit} and \eqref{equ:CKMfitter}, one gets
\begin{align}
\text{UTfit: }\quad\mathcal{B}(\kpn) &= \left(8.64^{+0.54}_{-0.53}\right)\times 10^{-11},\\
\mathcal{B}(\klpn) &= (2.93\pm 0.25)\times 10^{-11},\\
\text{CKMfitter: }\quad\mathcal{B}(\kpn) &= \left(8.17^{+0.61}_{-0.71}\right)\times 10^{-11},\\
\mathcal{B}(\klpn) &= \left(2.65^{+0.29}_{-0.28}\right)\times 10^{-11}.
\end{align}

It is also interesting to compare the results in \eqref{bsmumufinal}, \eqref{bdmumufinal} with the most recent prediction in the SM \cite{Bobeth:2013uxa}, with which our SM results are in perfect agreement,\footnote{This is not surprising as these authors used the 
inclusive determination of $\vcb$ that is very close to the value determined by us.}
and with the 
most recent averages from the combined analysis of CMS and LHCb \cite{CMS:2014xfa} that read
\begin{align}\label{LHCb2}
\overline{\mathcal{B}}(B_{s}\to\mu^+\mu^-) &= (2.8^{+0.7}_{-0.6}) \times 10^{-9},\\
\mathcal{B}(B_{d}\to\mu^+\mu^-) &= (3.9^{+1.6}_{-1.4})\times 10^{-10}.
\end{align}
Note that the SM value of $\overline{\mathcal{B}}(B_{s}\to\mu^+\mu^-)$ is outside one sigma range of the experimental value.

\begin{figure}[t]
\centering%
\includegraphics[width=0.45\textwidth]{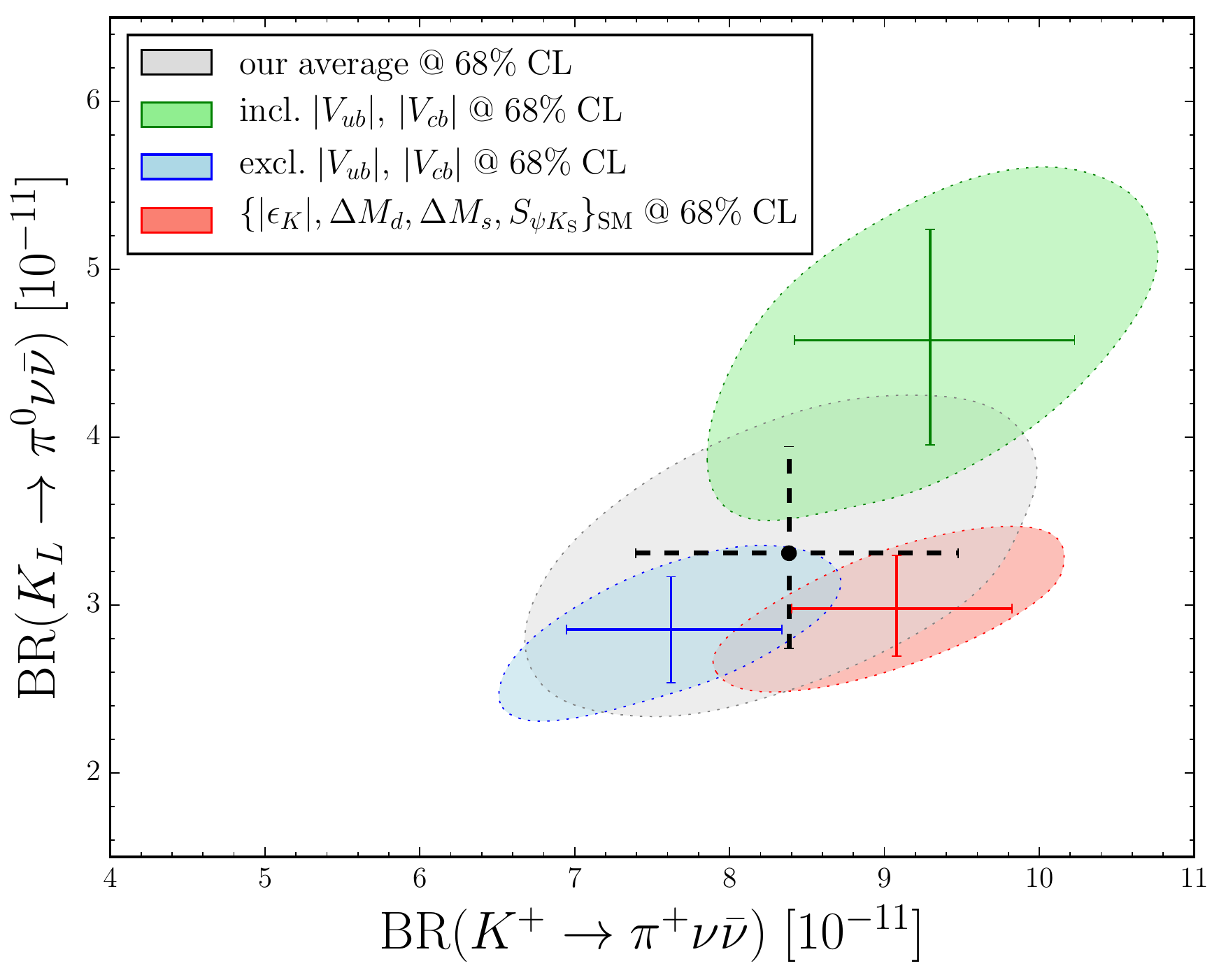}%
\includegraphics[width=0.455\textwidth]{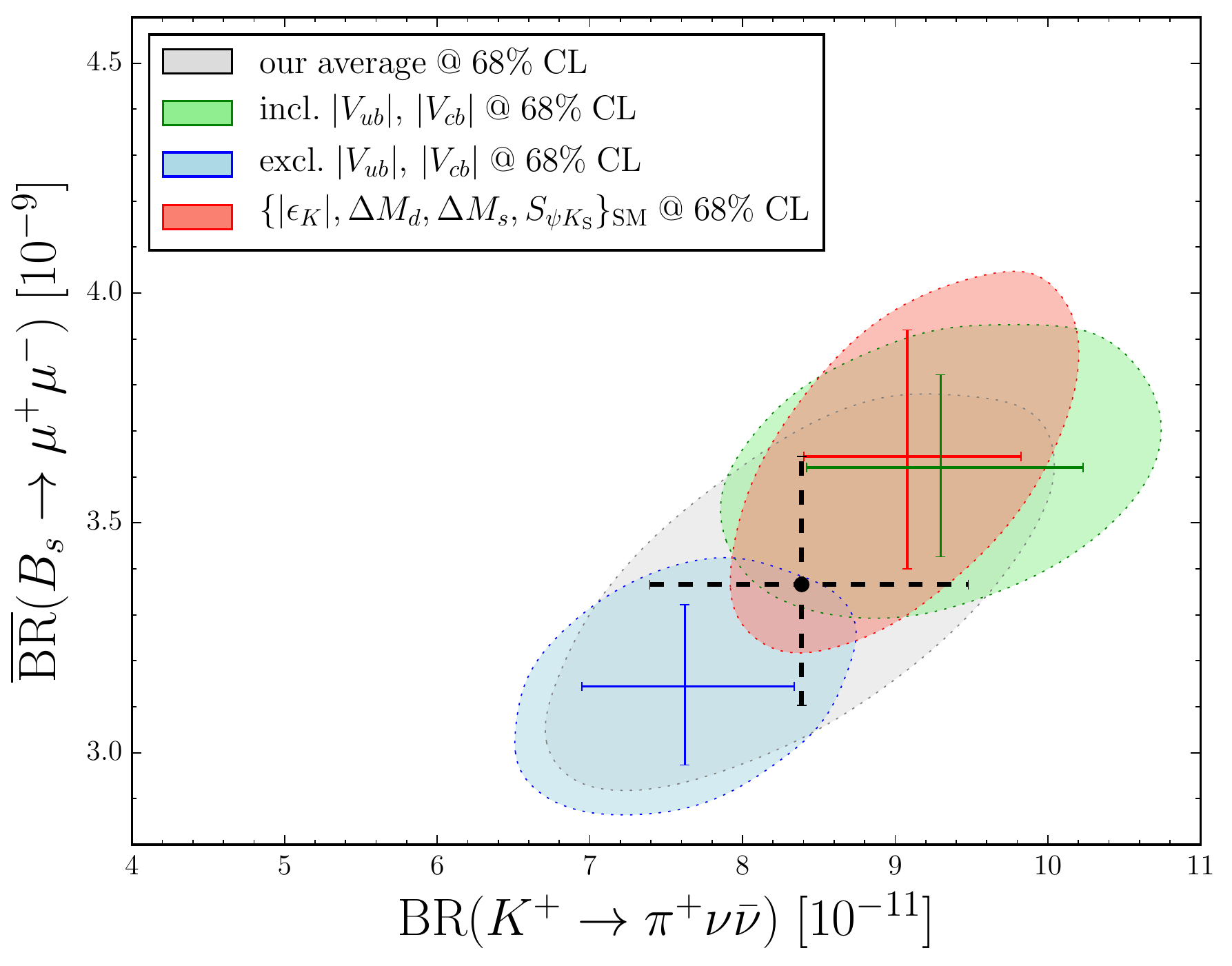}
\caption{\it Comparison of 68\% C.L. regions for $\mathcal{B}(\klpn)$ and $\overline{\mathcal{B}}(B_s\to\mu^+\mu^-)$ versus $\mathcal{B}(\kpn)$, using different inputs from both strategy A and B to fix the CKM matrix .
\label{fig:grandKnnPlots}}
\end{figure}

In Figure~\ref{fig:grandKnnPlots} the correlations of $\mathcal{B}(\klpn)$ and $\overline{\mathcal{B}}(B_s\to\mu^+\mu^-)$ versus $\mathcal{B}(\kpn)$ are shown, comparing the best result of strategy B, which includes all of the available inputs, with the inclusive, exclusive and average cases of strategy A.
We observe that the inclusive case of strategy A is very similar to strategy B for $\kpn$ and $\Bsmumu$, as both have little sensitivity to $\vub$, whereas $\klpn$, which has a stronger $\vub$ dependence, can differentiate them.
In both plots our average for $\vub$ and $\vcb$ is seen to also pick the  middle ground for these observables.

Evidently, the present experimental value for $\overline{\mathcal{B}}(B_{s}\to\mu^+\mu^-)$ in (\ref{LHCb2}) would favour the exclusive determination of $\vcb$ and a value 
of $\mathcal{B}(\kpn)$ in the ballpark of $7 \times 10^{-11}$ rather than 
 $9 \times 10^{-11}$. But then also the value of $|\varepsilon_K|$ would be 
below the data. It appears then that unless the experimental value for 
$\overline{\mathcal{B}}(B_{s}\to\mu^+\mu^-)$ moves up by $20\%$ in the 
coming years, the SM will face some tensions in this sector of flavour 
physics.

It is instructive to recall the following formula \cite{Buchalla:1998ba,D'Ambrosio:2001zh} that summarises the dependence of $\mathcal{B}(K^+ \rightarrow \pi^+ \nu \bar{\nu})$ on $R_t$, $\beta$ and $V_{cb}$:
\begin{align}
\label{AIACD}
\mathcal{B}(K^+ \rightarrow \pi^+ \nu \bar{\nu}) =\,
&\frac{\kappa_+}{\lambda^8}~\vcb^4 X(x_t)^2
\Bigg[\frac{R^2_t\sin^2\beta}{(1 - \lambda^2/2)^2}\notag\\
&+
\Big(1-\frac{\lambda^2}{2}\Big)^2\left(R_t\cos\beta +
\frac{\lambda^4P_c(X)}{\vcb^2X(x_t)}\right)^2\Bigg].
\end{align}
This can be considered as the fundamental formula for a correlation between 
$\mathcal{B}(\kpn)$, $\beta$ and any observable used to determine $R_t$, and 
is valid also in all models with MFV where $X(x_t)$ is replaced by a real function $X$. When this formula was proposed, it contained significant uncertainties 
in $R_t$ determined through $\Delta M_d/\Delta M_s$, in  $P_c(X)$ known only at 
NLO,  in $\kappa_+$ and in $\vcb$. The first three uncertainties have 
been significantly reduced since then. Moreover, the improved knowledge 
of the non-perturbative parameters entering $\varepsilon_K$ and $\Delta M_{s,d}$
allows now within the SM to determine $\vcb$ rather precisely. We stress  that in other models with MFV the latter determination will depend on the NP contributions to $\varepsilon_K$ and $\Delta M_{s,d}$ which modify the function $S$. An analysis of this issue is presented in  \cite{Buras:2013raa}.

Finally when $\gamma$ from tree-level decays will be precisely 
measured, $R_t$ will be determined solely by $\beta$ and $\gamma$,
\begin{equation}
R_t = \frac{\sin\gamma}{\sin(\beta+\gamma)},
\end{equation}
and the dependence of $\mathcal{B}(K^+\to\pi^+\nu\bar\nu)$ on $\gamma$ can be directly read off \eqref{AIACD}.


\boldmath
\section{The ratio $\epe$ in the Standard Model}\label{sec:4a}
\unboldmath

The ratio $\epe$  measures the size of direct CP
violation in $K_L\to\pi\pi$ 
relative to the indirect CP violation described by $\varepsilon_K$. 
In the SM $\varepsilon^\prime$ is governed by QCD penguins, but 
receives also an important destructively interfering contribution from electroweak penguins that is generally much more sensitive to NP than the QCD contribution.

The ratio $\epe$ is measured
to be \cite{Beringer:1900zz,Batley:2002gn,AlaviHarati:2002ye,Worcester:2009qt} 
\be\label{eprime}
\RE(\epe)=(16.5\pm 2.6)\times 10^{-4},
\ee
and the imaginary part of $\epe$ is negligible so that we 
will just write $\epe$ in all formulae below.

This result constitutes in principle a strong constraint on theory.
However, the difficulty in  making predictions for $\epe$ within the SM and 
its extensions is the strong cancellation of 
QCD penguin contributions and electroweak penguin contributions to
 this ratio. In the SM QCD penguins give a positive contribution, while the electroweak penguins a negative one.  In order to obtain a useful prediction for $\epe$ in the SM the corresponding hadronic parameters
$\bsi$ and $\bei$ have to  be known precisely. 

In the large $N$ limit one has $\bsi=\bei=1$ \cite{Buras:1985yx,Bardeen:1986vp,Buras:1987wc}. While the study of $1/N$ corrections in \cite{Hambye:1998sma} indicated that 
$\bei < 1$, no conclusive result has been obtained for $\bsi$. Fortunately, 
very recently significant progress on both $\bsi$ and $\bei$ has been made 
by lattice QCD simulations and in the context of the large $N$ approach. Indeed,
from the results of the RBC-UKQCD collaboration \cite{Bai:2015nea,Blum:2015ywa} one can extract 
\begin{align}\label{B8B6LAT0}
\bsi &= 0.57\pm 0.19, & B_8^{(3/2)} &= 0.76\pm 0.05\,, & {\rm (lattice~QCD)}
\end{align}
as shown in Appendix~\ref{app:epe} in the case of $\bei$, and in \cite{Buras:2015yba,Buras:2015xba} for $\bsi$.
On the other hand, the very recent analysis in the large-$N$ approach in
\cite{Buras:2015xba} allows to derive a conservative upper bound on both $\bsi$ and $\bei$, which reads
\begin{align}\label{NBOUND}
\bsi &\le \bei < 1. & &&(\mbox{\rm large-}N)
\end{align}
Moreover, one finds  $B_8^{(3/2)}(m_c)=0.80\pm 0.10$ in good agreement with 
(\ref{B8B6LAT0}). The result for $\bsi$ is less
precise but there is a strong indication that $\bsi < \bei$, also in agreement with
(\ref{B8B6LAT0}). We refer to \cite{Buras:2015xba} for further arguments 
why $\bsi$ is expected to be smaller than $\bei$.

The most recent analysis of $\epe$ has been given in \cite{Buras:2015yba}. 
Using the results in (\ref{B8B6LAT0}) and determining the remaining contributions to $\epe$ by imposing the agreement of the SM with CP-conserving data one 
finds\footnote{To this end ${\rm Im}\,\lambda_t = (1.4\pm0.1)\times 10^{-4}$ has been used.} \cite{Buras:2015yba}
\be\label{LBGJJ}
 \RE(\epe) = (1.9 \pm 4.5) \times 10^{-4} \,,
\ee
significantly below the experimental value in (\ref{eprime}). This result 
differs by  roughly $3\sigma$ from the data, but, as stressed in 
\cite{Buras:2015yba}, larger values can be obtained if only 
the absolute large $N$ upper bound on both parameters in (\ref{NBOUND}) 
is used. Yet, as found there and confirmed here by us, even with 
more generous values of ${\rm Im}\lambda_t$ the SM has serious difficulty 
in describing the data for $\epe$. 

In spite of this it is of interest to study the correlation of  $\epe$  
with $\klpn$ in the SM as this correlation has been already studied in various extensions of the SM  \cite{Buras:1998ed,Buras:1999da,Blanke:2007wr,Bauer:2009cf,Buras:2014sba,Buras:2015yca,Blanke:2015wba}. Within the SM 
this correlation depends only on the values of $\bsi$ and $\bei$ and the CKM parameters which we determined in the previous sections using strategies A and B. 

All the relevant  details on $\epe$ within the SM including the relevant references are given in \cite{Buras:2015yba}. Here we collect only the 
relevant information necessary to perform the numerical analysis. The 
basic analytic formula for $\epe$ reads \cite{Buras:2015yba}
\be 
\left(\frac{\varepsilon'}{\varepsilon}\right)_{\rm SM}= \IM\lambda_t
\cdot F_{\varepsilon'}(x_t),
\label{epeth}
\ee
where 
\be
F_{\varepsilon'}(x_t) =P_0 + P_X \, X_0(x_t) + 
P_Y \, Y_0(x_t) + P_Z \, Z_0(x_t)+ P_E \, E_0(x_t)~,
\label{FE}
\ee
with the first term dominated by QCD-penguin contributions, the next three 
terms by electroweak penguin contributions, and the last term being 
totally negligible. The $x_t$ dependent functions have been 
collected in Appendix~\ref{app:epe}.

The coefficients $P_i$ are given in terms of the non-perturbative parameters
$R_6$ and $R_8$ defined in (\ref{RS}) as follows:
\begin{equation}
P_i = r_i^{(0)} + 
r_i^{(6)} R_6 + r_i^{(8)} R_8 \,.
\label{eq:pbePi}
\end{equation}
The coefficients $r_i^{(0)}$, $r_i^{(6)}$ and $r_i^{(8)}$ comprise
information on the Wilson-coefficient functions of the $\Delta S=1$ weak
effective Hamiltonian at NLO. Their numerical values are given in the NDR renormalisation scheme for $\mu=m_c$ and three  values of $\alpha_s(M_Z)$
in Table~\ref{tab:pbendr} in Appendix~\ref{app:epe}.

The parameters
$R_6$ and $R_8$ are directly related to the parameters $\bsi$ and $\bei$ 
representing the hadronic matrix elements of $Q_6$ and $Q_8$, respectively. 
They are defined as
\begin{align}\label{RS}
R_6&\equiv \bsi\left[ \frac{114.54\mev}{m_s(m_c)+m_d(m_c)} \right]^2,\\
R_8&\equiv \bei\left[ \frac{114.54\mev}{m_s(m_c)+m_d(m_c)} \right]^2.
\end{align}
 We stress that both $\bsi$ and $\bei$ 
depend very weakly on the renormalisation scale \cite{Buras:1993dy}.

\begin{figure}[t]
\centering%
\includegraphics[width=0.6\textwidth]{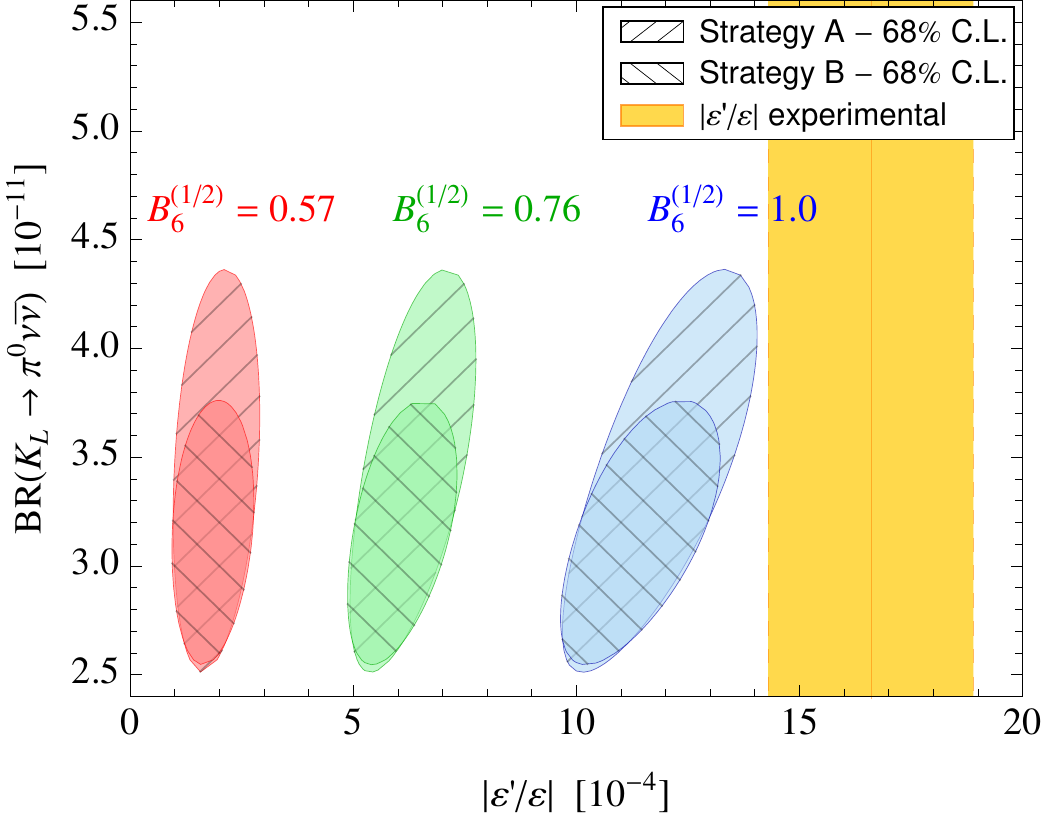}
\caption{\it Correlation of $\mathcal{B}(\klpn)$ versus $|\varepsilon'/\varepsilon|$ for fixed values of $\bsi$ = 0.57 (red), 0.76 (green), and 1.00 (blue). The hatched regions correspond to a 68\% C.L. resulting from the uncertainties on all other inputs, for strategy A using our average values of $\vub$ and $\vcb$, and strategy B. The yellow band shows the experimental result at $1\sigma$.
\label{fig:EpsPrimePlot}}
\end{figure}

In Figure~\ref{fig:EpsPrimePlot} we show the correlation between $\epe$ and $\klpn$ in the SM. The central value from the RBC-UKQCD collaboration in (\ref{B8B6LAT0}) has been used for $\bei$. The different colours correspond to different choices of the parameter $\bsi$:
\begin{align}
&& \bsi&=1.0 & &{(\text{blue})}, && \label{AJB1}\\
&& \bsi&=0.76 & &{(\text{green})}, && \label{AJB2}\\
&& \bsi&=0.57 & &{(\text{red})}\,. && \label{AJB3}
\end{align}
The first choice is motivated by the upper limit from large $N$ approach 
in (\ref{NBOUND}), although the bound $\bsi < \bei$ is violated, and gives an idea of the largest possible values of $\epe$ attainable in the SM.
The second choice assumes that $\bsi=\bei$ is saturating the previous bound.
Finally, the third choice uses the
central values (\ref{B8B6LAT0}) from the RBC-UKQCD collaboration for both $\bsi$ and $\bei$. We observe that even for the first choice of $\bsi$ and 
$\bei$ the ratio $\epe$ in the SM is below the data, and only for the largest 
values of ${\rm Im}\lambda_t$ it is within $2\sigma$ from the central experimental value. For such values also the branching ratio for $\klpn$ is largest.


\section{Summary and outlook}\label{sec:concl}

In this paper we have performed a new analysis of the rare decays $\kpn$ and 
$\klpn$ within the SM. The prime motivations for this study were:
\begin{itemize}
\item
The start of the NA62 experiment that should in the coming years reach a 
precision of $10\%$ relative to the SM prediction for $\mathcal{B}(\kpn)$.
\item
    The soon to improve value of $\xi$, for which preliminary error estimates are already given in \cite{Bouchard:2014eea}, which will allow 
a much more precise determination of the elements of the CKM matrix, in 
particular of the angle $\gamma$, $\vub$ and $\vcb$, without the use of 
present tree-level determinations of these parameters that are presently 
subject to significant uncertainties.
\item
The observation of the correlation between $\mathcal{B}(\kpn)$, $\mathcal{B}(B_s\to\mu^+\mu^-)$ and $\gamma$ within the SM that only weakly depends on $\vcb$. 
This correlation should be of interest in particular for CERN experimentalists 
who in the coming years will significantly improve the measurements on 
these three quantities. 
\end{itemize}

Our main results are illustrated with several
plots in Sections~\ref{sec:A} and \ref{sec:B}.
Our analysis demonstrates that in the coming years the SM will  undergo 
an unprecedented test due to the measurements of the rates for the decays 
$\kpn$ and $B_{s}\to\mu^+\mu^-$ and improved determinations of the CKM parameters either through the strategies A or B, accompanied by improved lattice QCD calculations of the relevant non-perturbative parameters. Around 2020 these studies will 
be enriched through precise measurements of the rates for $\klpn$, 
$B_{d}\to\mu^+\mu^-$  and $B_d\to K(K^*)\nu\bar\nu$ \cite{Buras:2014fpa}. 

{Also improved knowledge of the parameters $\bsi$ and $\bei$, accompanied with improved 
values of CKM parameters, will allow a more precise prediction for the important 
ratio $\epe$. Calculating this ratio using strategies A and B we find, in 
accordance with the recent analysis in \cite{Buras:2015yba}, that the 
SM prediction for $\epe$ is significantly below the data leaving large room 
for NP contributions. A recent analysis of $\epe$ in simplified NP models 
has shown which NP models could move the theory prediction for $\epe$ to agree 
with data \cite{Buras:2015yca}. Needless to say, in order to be sure that 
the SM indeed fails in the description of data, a big effort in clarifying various uncertainties will be required, as discussed in \cite{Buras:2015yba}.}

It should be observed that the agreement of the SM prediction for 
$\overline{\mathcal{B}}(B_s\to\mu^+\mu^-)$ with the data can be significantly improved 
by lowering $\vcb$ to the values in the ballpark of its present exclusive determinations using lattice QCD form factors. But then automatically $\varepsilon_K$ 
is found significantly below the data. Interestingly in this case 
 $\mathcal{B}(\kpn)$ is also predicted to be in the ballpark of $7\times 10^{-11}$, that is more than a factor of two below its present experimental average. 
No doubt, the coming years will be exceptional for quark flavour physics.


\section*{Acknowledgements}
We thank Aida X. El-Khadra and Andreas S. Kronfeld for illuminating and informative discussions about their lattice calculations of $\Delta B=2$ matrix elements  and Tadeusz Jankowski and Chris Sachrajda for the insight in their calculation
of $\Delta I=3/2$ $K\to\pi\pi$ matrix elements.
This research was done and financed in the context of the ERC Advanced Grant project ``FLAVOUR''(267104) and was partially
supported by the DFG cluster
of excellence ``Origin and Structure of the Universe''.


\appendix
\boldmath
\section{Expression for $X_t$}\label{app:X}
\unboldmath

The loop function $X_t$ of \eqref{XSM} can be written as
\begin{equation}
X(x_t) = X_0(x_t) + \frac{\alpha_s}{4\pi}X_1(x_t) + \frac{\alpha}{4\pi}X_{\rm EW}(x_t),
\end{equation}
where $X_0$ is the leading order result, and $X_1$, $X_{\rm EW}$ are the NLO QCD and EW corrections, respectively. The coupling constants $\alpha_s$ and $\alpha$, as well as the parameter $x_t = m_t^2/m_W^2 = 2 y_t^2/g_2^2$, have to be evaluated at a given renormalisation scale $\mu\sim\mathcal{O}(M_t)$.

The LO expression is 
\begin{equation}\label{X01}
X_0(x_t) = \frac{x_t}{8}\left[\frac{x_t+2}{x_t-1} + \frac{3x_t-6}{(x_t-1)^2}\log x_t\right].
\end{equation}
The NLO QCD correction \cite{Buchalla:1993bv,Misiak:1999yg,Buchalla:1998ba} reads, in the $\overline{\text{MS}}$ scheme,
\begin{equation}\begin{aligned}
X_1(x_t) &= -\frac{29x_t - x_t^2 - 4x_t^3}{3(1-x_t)^2} - \frac{x_t + 9x_t^2 - x_t^3 - x_t^4}{(1-x_t)^3}\log x_t\\
&+ \frac{8x_t + 4x_t^2 + x_t^3 - x_t^4}{2(1-x_t)^3}\log^2 x_t - \frac{4x_t - x_t^3}{(1-x_t)^2}{\rm Li}_2(1-x_t)\\
&+ 8x_t\frac{\partial X_0}{\partial x_t}\log\frac{\mu^2}{M_W^2},
\end{aligned}\end{equation}
where $\mu$ is the renormalisation scale.
The 2-loop EW correction $X_{\rm EW}$ has been calculated in \cite{Brod:2010hi}, but no explicit result has been presented. Approximate formulae, one of which accurate to more than 0.05\%, as well as a plot of the contribution $(\alpha/4\pi) X_{\rm EW}$ can however be found in \cite{Brod:2010hi}.

\begin{figure}
\centering%
\includegraphics[width=0.56\textwidth]{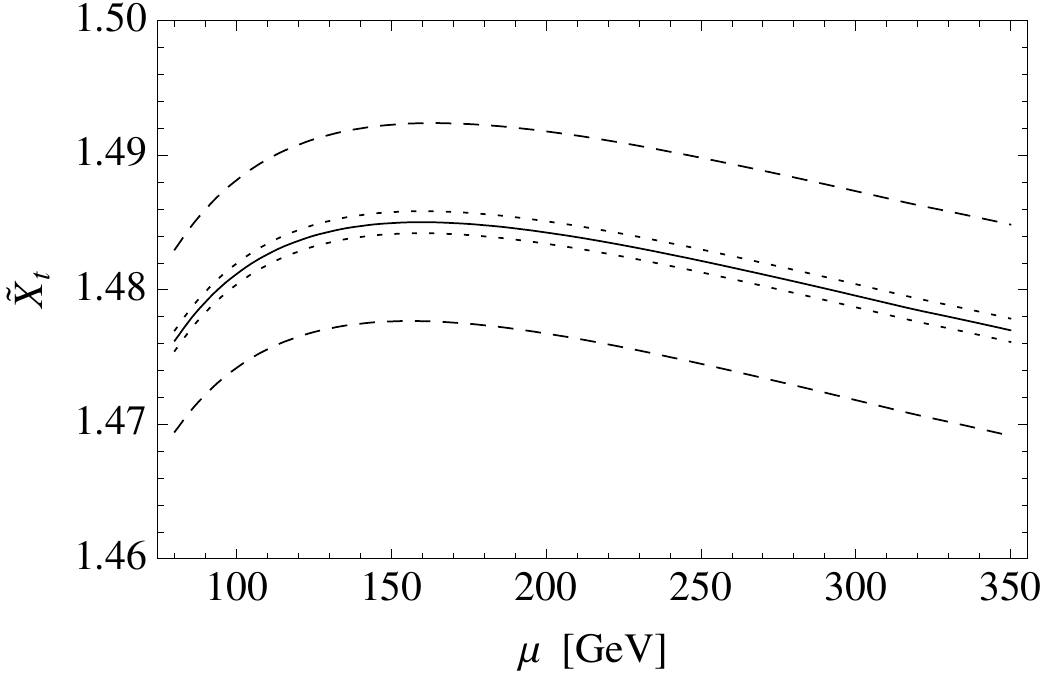}\hfill%
\includegraphics[width=0.36\textwidth]{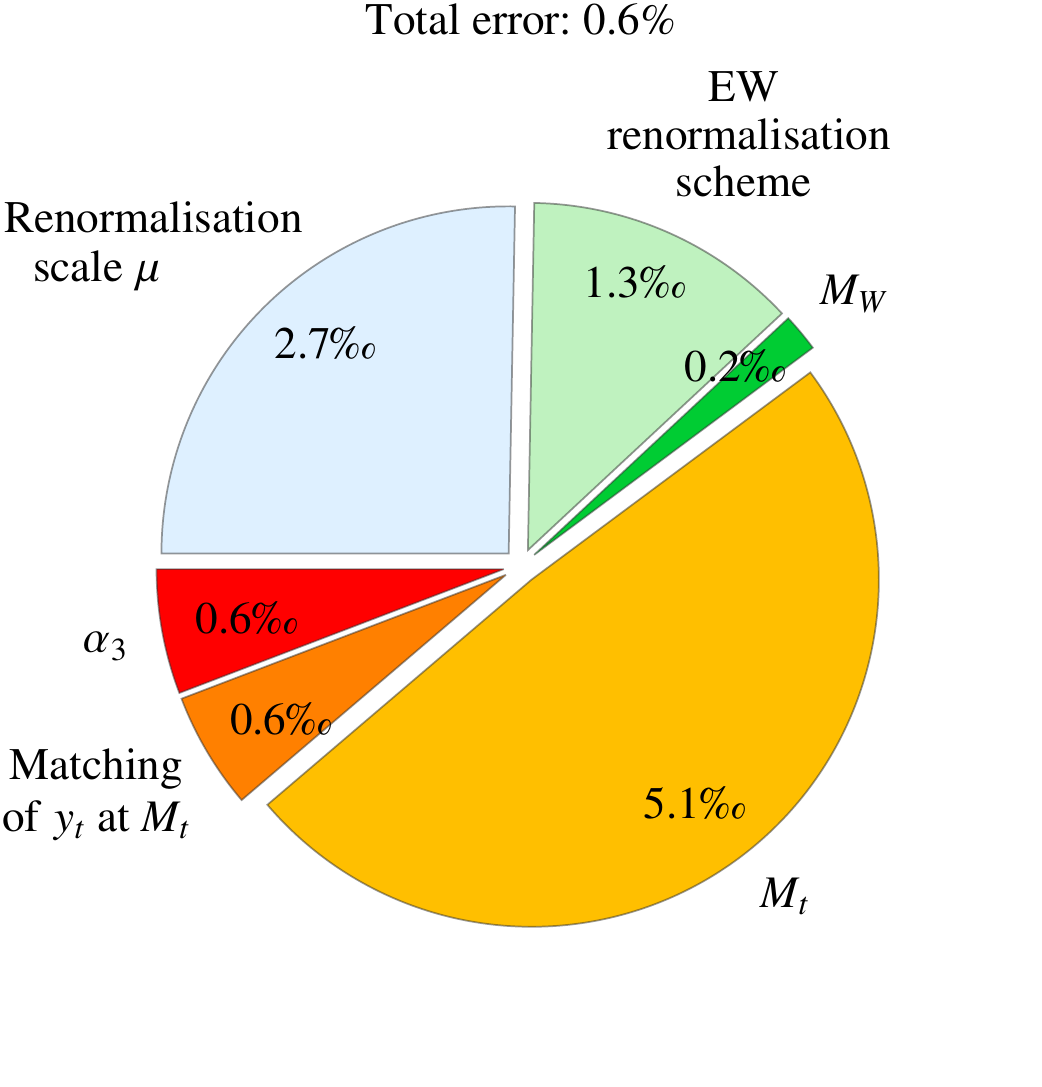}
\caption{Left: Renormalisation scale dependence of the quantity $\tilde X(\mu)$. The dashed lines show the uncertainty due to the error on the measured pole top mass $M_t$, while the dotted lines correspond to the theoretical error on the $\overline{\text{MS}}$ top Yukawa coupling $y_t$ due to higher orders in the matching at the weak scale. Right: different sources of error affecting $X_t$.\label{fig:Xt}}
\end{figure}

The left panel of Figure~\ref{fig:Xt} shows a plot of the scale- and scheme-independent quantity
\begin{equation}\label{Xtilde}
\tilde X(\mu) = \frac{\alpha(\mu)}{\sin^2\theta_w(\mu)}\frac{\sin^2\theta_w(M_Z)}{\alpha(M_Z)}X(x_t(\mu),\mu)
\end{equation}
as a function of the renormalisation scale $\mu$, together with the 1$\sigma$ bands corresponding to the theoretical error in the matching of the top Yukawa coupling $y_t$ at the weak scale, and the experimental error on the top mass $M_t$. The $\overline{\text{MS}}$ couplings at full NNLO -- 2-loop matching at the weak scale (3-loop QCD for $\alpha_s$ and $y_t$) and 3-loop running (4-loop QCD for $\alpha_s$) -- as determined in \cite{Buttazzo:2013uya} have been used. The remaining scale dependence shown in the figure comes from higher order corrections, mainly from QCD, and accounts for an error of 0.004 on $X_t$. An additional error of 0.002 comes from the ambiguity in the choice of the renormalisation scheme for the EW prefactor, as shown in \cite{Brod:2010hi}. A comparison of the different errors contributing to $X_t$ is shown in the right panel of Figure~\ref{fig:Xt}. The experimental error on the top quark pole mass $M_t$ is by far the dominant contribution at present.


\boldmath
\section{Expression for $P_c(X)$}\label{app:Pc}
\unboldmath

An approximate formula for $P^{\rm SD}_c(X)$ taken from~\cite{Brod:2008ss} reads
\begin{align}
\begin{split}
    P^{\rm SD}_c(X) &= 0.38049 \left(\frac{m_c(m_c)}{1.30~\text{GeV}}\right)^{0.5081} \left(\frac{\alpha_s(M_Z)}{0.1176}\right)^{1.0192}\left(1+\sum_{i,j}\kappa_{ij}L_{m_c}^i L_{\alpha_s}^j\right) \\
  &\pm 0.008707\left(\frac{m_c(m_c)}{1.30~\text{GeV}}\right)^{0.5276}\left(\frac{\alpha_s(M_Z)}{0.1176}\right)^{1.8970}\left(1+\sum_{i,j}\epsilon_{ij}L_{m_c}^i L_{\alpha_s}^j\right),
\end{split}
\end{align}
where
\begin{align}
 &L_{m_c} = \ln\left(\frac{m_c(m_c)}{1.30~\text{GeV}}\right)\,,\qquad L_{\alpha_s} = \ln \left(\frac{\alpha_s(M_Z)}{0.1176}\right)
\end{align}
and
\begin{align}
 &\kappa_{10} = 1.6624,\quad \kappa_{01} = -2.3537,\quad \kappa_{11} = -1.5862,\quad \kappa_{20} =  1.5036,\quad \kappa_{02} =  -4.3477,\nonumber\\
 &\epsilon_{10} = -0.3537,\quad \epsilon_{01} =  0.6003,\quad\epsilon_{11} =  -4.7652,\quad\epsilon_{20} = 1.0253,\quad\epsilon_{02} =  0.8866.
\end{align}

\boldmath
\section{More details on $\epe$}\label{app:epe}
\unboldmath

The basic one-loop functions entering (\ref{FE}) are given  by (\ref{X01}) and
\begin{equation}\label{Y0}
Y_0(x_t)={\frac{x_t}{8}}\left[{\frac{x_t -4}{x_t-1}} 
+ {\frac{3 x_t}{(x_t -1)^2}} \ln x_t\right],
\end{equation}
\bea\label{Z0}
Z_0(x_t)~\!\!\!\!&=&\!\!\!\!~-\,{\frac{1}{9}}\ln x_t + 
{\frac{18x_t^4-163x_t^3 + 259x_t^2-108x_t}{144 (x_t-1)^3}}+
\nonumber\\ 
&&\!\!\!\!+\,{\frac{32x_t^4-38x_t^3-15x_t^2+18x_t}{72(x_t-1)^4}}\ln x_t
\eea
\begin{equation}\label{E0}
E_0(x_t)=-\,{\frac{2}{3}}\ln x_t+{\frac{x_t^2(15-16x_t+4x_t^2)}{6(1-x_t)^4}}
\ln x_t+{\frac{x_t(18-11x_t-x_t^2)}{12(1-x_t)^3}} ~,
\end{equation}
where $x_t=m^2_t/M_W^2$.

The coefficients $r_i^{(0)}$, $r_i^{(6)}$ and $r_i^{(8)}$ entering (\ref{eq:pbePi}) are given 
in the NDR renormalisation scheme for $\mu=m_c$ and three  values of $\alpha_s(M_Z)$
in Table~\ref{tab:pbendr}.

The parameters $B_6^{(1/2)}$ and $B_8^{3/2}$ are related to the hadronic matrix elements $Q_6$ and $Q_8$ as follows
\be\label{eq:Q60}
\langle Q_6(\mu) \rangle_0=-\,4 
\left[ \frac{m_{\rm K}^2}{m_s(\mu) + m_d(\mu)}\right]^2 (F_K-F_\pi)
\,B_6^{(1/2)}\,,
\ee
\be\label{eq:Q82}
\langle Q_8(\mu) \rangle_2=\sqrt{2}
\left[ \frac{m_{\rm K}^2}{m_s(\mu) + m_d(\mu)}\right]^2 F_\pi
\,B_8^{(3/2)}\,.
\ee
It should be emphasised that the overall factor in these expressions depends on the normalisation of the amplitudes $A_{0,2}$. The matrix elements given above correspond to the normalisation used in \cite{Cirigliano:2011ny,Buras:2014maa,Buras:2014sba}. On the other hand the RBC-UKQCD collaboration \cite{Blum:2012uk,Blum:2015ywa} uses a different normalisation adopted in \cite{Buras:1993dy}. 
By comparing (\ref{eq:Q60}) and (\ref{eq:Q82}) with Eqs. (5.10) and (5.18) of the latter paper we find that the matrix elements in  \cite{Buras:1993dy} have 
and additional factor of $\sqrt{3/2}$. While $\epe$ clearly does not
depend on this difference, it is crucial to take it into account when extracting the value of $B_8^{(3/2)}$ from the results obtained by  RBC-UKQCD collaboration. To this end 
we use Eq. (30) for $A_2$ in \cite{Blum:2012uk}, adjust to our normalisation, and compare to $A_2$ expressed in terms of
$\langle Q_8(\mu) \rangle_2$ in (\ref{eq:Q82}). This allows us to related $\langle Q_8(\mu) \rangle_2$ to the hadronic matrix element $\mathcal{M}^{\overline{\text{MS}}-\text{NDR}}_{(8,8)_\text{mix}}$  used in \cite{Blum:2012uk,Blum:2015ywa}:
\be
\langle Q_8(\mu) \rangle_2=\frac{1}{3\sqrt{2}}\mathcal{M}^{\overline{\text{MS}}-\text{NDR}}_{(8,8)_\text{mix}}
\ee
In this manner we find
\be\label{eq:B8LAT}
B_8^{(3/2)}(\mu)=\frac{1}{6 F_\pi} \left[\frac{m_s(\mu) + m_d(\mu)}{m_{\rm K}^2}\right]^2 \mathcal{M}^{\overline{\text{MS}}-\text{NDR}}_{(8,8)_\text{mix}} (\mu)\,.
\ee
The $\mu$ dependence of $\mathcal{M}^{\overline{\text{MS}}-\text{NDR}}_{(8,8)_\text{mix}} (\mu)$ is practically cancelled by the one of quark masses so that $\bei$ is 
practically $\mu$-independent.
In particular in the 
$\overline{\text{MS}}$--NDR scheme the $\mu$-dependence is very weak 
\cite{Buras:1993dy}.

Using the QCD lattice value from \cite{Blum:2015ywa}\footnote{We thank Tadeusz Jankowski for translating the result in  \cite{Blum:2015ywa} into the 
$\overline{\text{MS}}-\text{NDR}$ scheme.}
\be
\mathcal{M}^{\overline{\text{MS}}-\text{NDR}}_{(8,8)_\text{mix}} (3\gev) =4.55\pm 0.27,
\ee
together with the light quark mass values~\cite{Aoki:2013ldr}
\be
m_s(2\gev)=(93.8\pm2.4) \mev, \qquad
m_d(2\gev)=(4.68\pm0.16)\mev,
\ee
we find 
\be
B_8^{(3/2)}(3\gev)=0.75\pm 0.05, \qquad B_8^{(3/2)}(m_c)=0.76\pm 0.05.
\ee

\begin{table}[!tb]
\begin{center}
\begin{tabular}{|c||c|c|c||c|c|c||c|c|c|}
\hline
&\multicolumn{3}{c||}{$\alpha_s(M_Z)= 0.1179$}&
 \multicolumn{3}{c||}{$\alpha_s(M_Z)= 0.1185$} &
  \multicolumn{3}{c|}{$\alpha_s(M_Z)= 0.1191$} \\
  \hline
$i$ & $r_i^{(0)}$ & $r_i^{(6)}$ & $r_i^{(8)}$ &
      $r_i^{(0)}$ & $r_i^{(6)}$ & $r_i^{(8)}$ &
       $r_i^{(0)}$ & $r_i^{(6)}$ & $r_i^{(8)}$ \\
      \hline
0 &    -3.392 & 15.293 & 1.271 &
       -3.421 & 15.624 & 1.231 &
       -3.451 & 15.967 & 1.191 \\
$X_0$ & 0.655 &  0.029 & 0. &
        0.655 &  0.030 & 0. &
        0.655 &  0.031 & 0. \\
$Y_0$ & 0.451 &  0.114 & 0. &
        0.449 &  0.116 & 0. &
        0.447 &  0.118 & 0. \\
$Z_0$ & 0.406 & -0.022 & -13.435 &
        0.420 & -0.022 & -13.649 &
        0.435 & -0.023 & -13.872 \\
$E_0$ & 0.229 & -1.760 & 0.652 &
        0.228 & -1.788 & 0.665 &
        0.226 & -1.816 & 0.678 \\
\hline
\end{tabular}
\end{center}
\caption{\it The coefficients $r_i^{(0)}$, $r_i^{(6)}$ and $r_i^{(8)}$ of
formula (\ref{eq:pbePi}) in the  NDR-${\rm \overline{MS}}$  scheme for three values of $\alpha_s(M_Z)$. From \cite{Buras:2015yba}.
\label{tab:pbendr}}
\end{table}


\renewcommand{\refname}{R\lowercase{eferences}}

\addcontentsline{toc}{section}{References}

\bibliographystyle{JHEP}
\bibliography{allrefs}
\end{document}